\documentclass[aps,prl,preprint,superscriptaddress]{revtex4-1}

\usepackage{graphicx}

\usepackage{bm}
\usepackage{color}
\usepackage{url}
\usepackage{tikz}
\usetikzlibrary{calc}
\usepackage{pgfplots}
\usepackage{amsmath,amssymb}
\usepackage{verbatim}
\usepackage{physics}
\usepackage{xcolor}
\usepackage{mwe}
\usepackage{tikz}
\usetikzlibrary{calc}
\usepackage{pgfplots}

\newcommand{\bea}{\begin{eqnarray}}
\newcommand{\eea}{\end{eqnarray}}

\newcommand{\br}{\mathbf{r}}

\newcommand{\be}{\begin{equation}}
\newcommand{\ee}{\end{equation}}
\newcommand{\bk}{\mathbf{k}}

\newcommand{\beal}{\begin{align}}
\newcommand{\eeal}{\end{align}}

\newcommand{\I}{\text{i}}

\begin{document}

\title{Anomalous Kerr effect in SrRuO$_3$ thin films}
\author{F. Michael Bartram}
\thanks{F.M.B., S.S. and Z.L. contributed equally to this work.}
\affiliation{%
Department of Physics, University of Toronto, Toronto, Ontario M5S 1A7, Canada
}%
\affiliation{%
State Key Laboratory of Low Dimensional Quantum Physics, Department of Physics, Tsinghua University, Beijing, 100084, China
}%

\author{Sopheak Sorn}
\thanks{F.M.B., S.S. and Z.L. contributed equally to this work.}
\affiliation{%
Department of Physics, University of Toronto, Toronto, Ontario M5S 1A7, Canada
}%

\author{Zhuolu Li}
\thanks{F.M.B., S.S. and Z.L. contributed equally to this work.}
\affiliation{%
State Key Laboratory of Low Dimensional Quantum Physics, Department of Physics, Tsinghua University, Beijing, 100084, China
}%

\author{Kyle Hwangbo}
\affiliation{%
Department of Physics, University of Toronto, Toronto, Ontario M5S 1A7, Canada
}%

\author{Shengchun Shen}
\affiliation{%
State Key Laboratory of Low Dimensional Quantum Physics, Department of Physics, Tsinghua University, Beijing, 100084, China
}%

\author{Felix Frontini}
\affiliation{%
Department of Physics, University of Toronto, Toronto, Ontario M5S 1A7, Canada
}%

\author{Liqun He}
\affiliation{%
Department of Physics, University of Toronto, Toronto, Ontario M5S 1A7, Canada
}%

\author{Pu Yu}
\email{Corresponding author. yupu@mail.tsinghua.edu.cn}
\affiliation{%
State Key Laboratory of Low Dimensional Quantum Physics, Department of Physics, Tsinghua University, Beijing, 100084, China
}%
\affiliation{%
Frontier Science Center for Quantum Information, Beijing 100084, China}%
\affiliation{%
RIKEN Center for Emergent Matter Science (CEMS), Wako 351-198, Japan}%

\author{Arun Paramekanti}
\email{Corresponding author. arunp@physics.utoronto.ca}
\affiliation{%
Department of Physics, University of Toronto, Toronto, Ontario M5S 1A7, Canada
}%

\author{Luyi Yang}
\email{Corresponding author. luyi-yang@mail.tsinghua.edu.cn}
\affiliation{%
Department of Physics, University of Toronto, Toronto, Ontario M5S 1A7, Canada
}%
\affiliation{%
State Key Laboratory of Low Dimensional Quantum Physics, Department of Physics, Tsinghua University, Beijing, 100084, China
}%
\affiliation{%
Frontier Science Center for Quantum Information, Beijing 100084, China}%

\date{\today}

\begin{abstract}
We study the magneto-optical Kerr effect (MOKE) in SrRuO$_3$ thin films, 
uncovering wide regimes of wavelength,
temperature, and magnetic field where the Kerr rotation is not simply proportional to the magnetization but instead displays two-component behavior. One component of the MOKE signal tracks the average magnetization, while the
second ``anomalous" component bears a resemblance to anomalies in the Hall resistivity which have been previously reported in skyrmion
materials. We present a theory showing that the MOKE anomalies arise from the non-monotonic relation between the Kerr angle and 
the magnetization, when we average over magnetic domains which proliferate near the coercive field. Our results suggest
that inhomogeneous domain formation, rather than skyrmions, may provide a common origin for the observed MOKE and Hall resistivity anomalies.
\end{abstract}


\maketitle

In magnetic solids, the nontrivial momentum space structure of Bloch bands \cite{TKNN,classification1, classification2, classification3}
reveals itself via the intrinsic anomalous Hall effect which arises due to spin-orbit coupling (SOC) and the band Berry curvature  \cite{Nagaosa2010}. 
Remarkably, three-dimensional (3D) magnetic solids can host topological
monopole-like singularities of the Berry curvature at Weyl nodes in the dispersion \cite{Fang2003, Ashvin2011, Mn3Sn1, Mn3Sn2}. 
These Weyl nodes can contribute significantly to $\rho_{xy}$ as seen in metallic antiferromagnets like Mn$_3$X (X \!=\! Sn, Ge) \cite{Mn3Sn1, Mn3Sn2},
and in metallic ferromagnets such as SrRuO$_3$ where $\rho_{xy}$ exhibits an unusual non-monotonic temperature dependence as the Weyl nodes approach the Fermi energy via tuning magnetization or temperature 
\cite{Fang2003, Nagaosa2010, Burkov2013}.

Magnetic solids can also harbor real-space topological textures called skyrmions. Skyrmions
imprint a real-space Berry curvature on conduction
electrons, inducing a so-called ``topological Hall effect'' (THE) \cite{THE1, THE2}. Skyrmion crystals and THE have been observed in various materials 
including MnSi \cite{THE1, THE2}, MnGe \cite{MnGe},  Ir/Fe/Co/Pt multilayers \cite{skx_roomtemperature,Legrandeaat0415,PhysRevLett.122.237201}, and Gd$_2$Pd$_3$Si \cite{Kurumaji2019}.  In recent work,
ultrathin epitaxial films of SrRuO$_3$ have been proposed as an oxide-based platform 
for hosting nanoscale skyrmions, stabilized by a strong chiral interfacial Dzyaloshinskii-Moriya (IDM) exchange \cite{Matsuno2016, Meng2019}. 

Experimental support for skyrmions in ultrathin SrRuO$_3$ films comes from the observation of
anomalous bump-like features in $\rho_{xy}$ near the coercive field while traversing a 
magnetic hysteresis loop \cite{Matsuno2016, Meng2019, Pang2017, Ohuchi2018, TWNoh_NatMat2018, Qin2019}, similar to THE anomalies
seen in other skyrmion materials.
However, alternative proposals suggest that the Hall anomalies in this ultrathin limit can arise from the complicated temperature dependence of the intrinsic $\rho_{xy}$,
together with atomic layer inhomogeneities in the ferromagnetic transition temperature $T_c$ or the coercive field 
\cite{AlternateExplanation, AlternateExplanation2,Noh2020, Malsch2020}. These alternative proposals rely on the momentum-space Berry curvature 
contribution to the intrinsic Hall effect, but the validity of such theories which simply add up the Hall response of two distinct regions remains unclear.

To assess the importance of momentum-space Berry curvature versus the role of skyrmions, and to further test the validity of these proposals, it is important to
examine distinct regimes using new experimental probes. In this Letter, we study the magneto-optical Kerr effect (MOKE)
in SrRuO$_3$ films. MOKE is a powerful, contactless,  and nondestructive technique with high 
sensitivity and submicrometer spatial resolution. It has been widely used to study electronic and magnetic properties in magnetic materials and devices \cite{QiuMOKE} and more recently in atomically-thin 2D van der Waals magnets \cite{XuCI3,ZhangCrGeTe}.

In contrast to previous work, our samples ($30$ nm to $200$ nm thick) are far from the ultrathin limit, so 
the IDM interaction, skyrmions, and atomic layer inhomogeneities, are not expected to play an important role. 
Remarkably, even in this regime, we discover bump-like 
anomalies in the MOKE signal over wide ranges of temperature, magnetic field, and frequencies, 
while the magnetization exhibits normal square-like hysteresis loops. 
Significantly, this observation contradicts the well-established lore that the polar Kerr rotation is proportional to the macroscopic magnetization in ferromagnetic thin films \cite{Matsuno2016,Ohuchi2018,QiuMOKE,kerrangle}.
We describe a controlled theory for the high frequency response,
showing that the MOKE anomalies can be semi-quantitatively captured by a combination of the
non-monotonic magnetization dependence of the 
Kerr angle and local averaging over magnetic domains. This experimental discovery of a Kerr anomaly, and its theoretical explanation, constitute the
key new significant results of our work.

{\it Experimental observations. --}
SrRuO$_3$ films were grown using pulsed-laser deposition onto both 
(LaAlO$_3$)$_{0.3}$(Sr$_2$TaAlO$_6$)$_{0.7}$ (LSAT) and SrTiO$_3$ (STO) 
substrates. In contrast to previous studies which explored ultrathin few unit-cell films, our SrRuO$_3$ films range in thickness from $30$ to $200$ nm, displaying ferromagnetic order below $T_c \! \sim \! 150$\,K \cite{LiNC2020}. Here, we focus on the case of an $88$\,nm thick film grown on LSAT;
see Supplemental Material (SM) \cite{suppmat} for data on other films.
We have measured the Kerr rotation in these films using a wavelength-tunable pulsed laser (repetition rate: 80 MHz) reduced to low-power (\textless1 mW) and focused weakly onto a $\sim$50 $\mu$m spot on the sample, with the reflected beam modulated by a photo-elastic modulator and passed through a Wollaston prism into a pair of balanced photodiodes, allowing measurement of both the real and imaginary parts of the Kerr angle. Measurements were done in a typical polar Kerr configuration, 
with the beam reflected at normal incidence and the external field applied in the out-of-plane direction.

\begin{figure}
 \includegraphics[width=0.7\columnwidth]{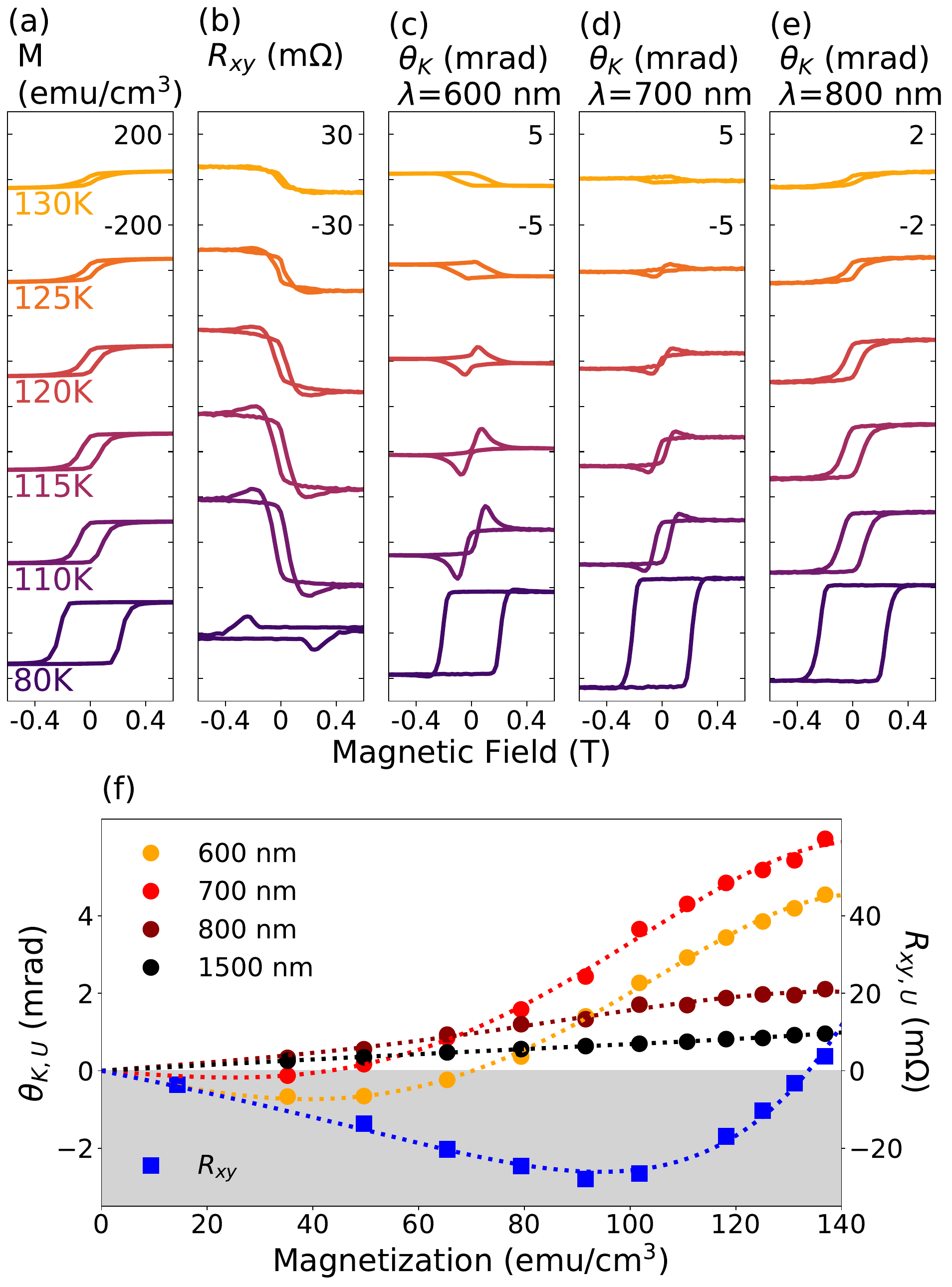}
 \caption{Comparison of (a) magnetization $M$, (b) Hall resistance $R_{xy}$, and Kerr rotation $\theta_K$ at normal incidence with (c) 600 nm, (d) 700 nm and (e) 800 nm lasers for an $88$\,nm thick SrRuO$_3$ sample on LSAT. Both $M$ and the $800$\,nm Kerr rotation data show a typical hysteresis loop. 
However, $R_{xy}$ and $\theta_K$ at 600 and 700 nm show additional bump features. The bump features in $\theta_K$ are of opposite sign when compared to those  
in the $R_{xy}$ measurements for the same field direction. 
Also shown in (f) are the values of $R_{xy}$ and $\theta_K$ at high fields, where the magnetization is uniform, which we define as $R_{xy,U}(T)$ and $\theta_{K,U}(T)$, plotted against the corresponding saturation magnetization $M(T)$. Dotted lines are guides to the eye. Note that an additional data set at 1500 nm not shown above is present here, the hysteresis loops here are very similar to those at 800 nm (see SM \cite{suppmat} for details). \label{fg:data}}
\end{figure}
 
Figure~\ref{fg:data} shows measurements of the Kerr rotation $\theta_K$ at normal incidence, Hall resistance $R_{xy}$, and magnetization $M$, at temperatures from $80$\,K 
to $130$\,K for an 88 nm thick SrRuO$_3$ sample on LSAT. We show and discuss only the real part of the Kerr rotation at $\lambda\!=\!$ 600, 700, and 800\,nm; 
the imaginary part, and additional data at $500$ and $1500$\,nm, can be found in the SM \cite{suppmat}. While the magnetization (a) has the shape of a 
typical hysteresis loop, both $R_{xy}$ (b), and the Kerr rotations at $\lambda\!=\!600$\,nm (c) and
$\lambda\!=\!700$\,nm (d), show a large additional bump-like contribution similar to the topological contribution
seen in systems hosting magnetic skyrmions. Such an additional contribution was
also observed in previous $R_{xy}$ measurements on few-unit cell ultrathin films, and was attributed to skyrmions \cite{Matsuno2016, Pang2017, Ohuchi2018, Qin2019, Meng2019, TWNoh_NatMat2018}. 
For longer wavelengths such as $\lambda\!=\!800$\,nm (e) and $\lambda\!=\!1500$\,nm (see SM), however, this additional contribution is no 
longer present in our samples, and $\theta_K$ tracks the magnetization curve.

In addition to these anomalies, the values of $R_{xy}$ and each $\theta_K$ at large fields, where the magnetization is uniform, change in very different way as the temperature (and therefore the value of this uniform magnetization) is varied.
To see this clearly, we show in (f) these values, which we define as $R_{xy,U}(T)$ and $\theta_{K,U}(T)$ plotted against the corresponding value of the saturation magnetization $M(T)$. (The subscript $U$ denotes ``uniform'' since the high field regime is expected to have a uniform magnetization across the sample.)
For the data at $\lambda\!=\!800$\,nm and  $\lambda\!=\!1500$\,nm, the result of this is approximately linear. Because at these wavelengths the hysteresis loops simply track the magnetization (with no anomalies), we conclude that $\theta_K \propto M$ across all temperatures and field values. For the $R_{xy}$, $\lambda\!=\!600$\,nm and  $\lambda\!=\!700$\,nm data, however, there is a clearly nonlinear relationship, including a zero crossing at some $M\textgreater0$ for each of them.

\begin{figure}[t]
 \includegraphics[width=\columnwidth]{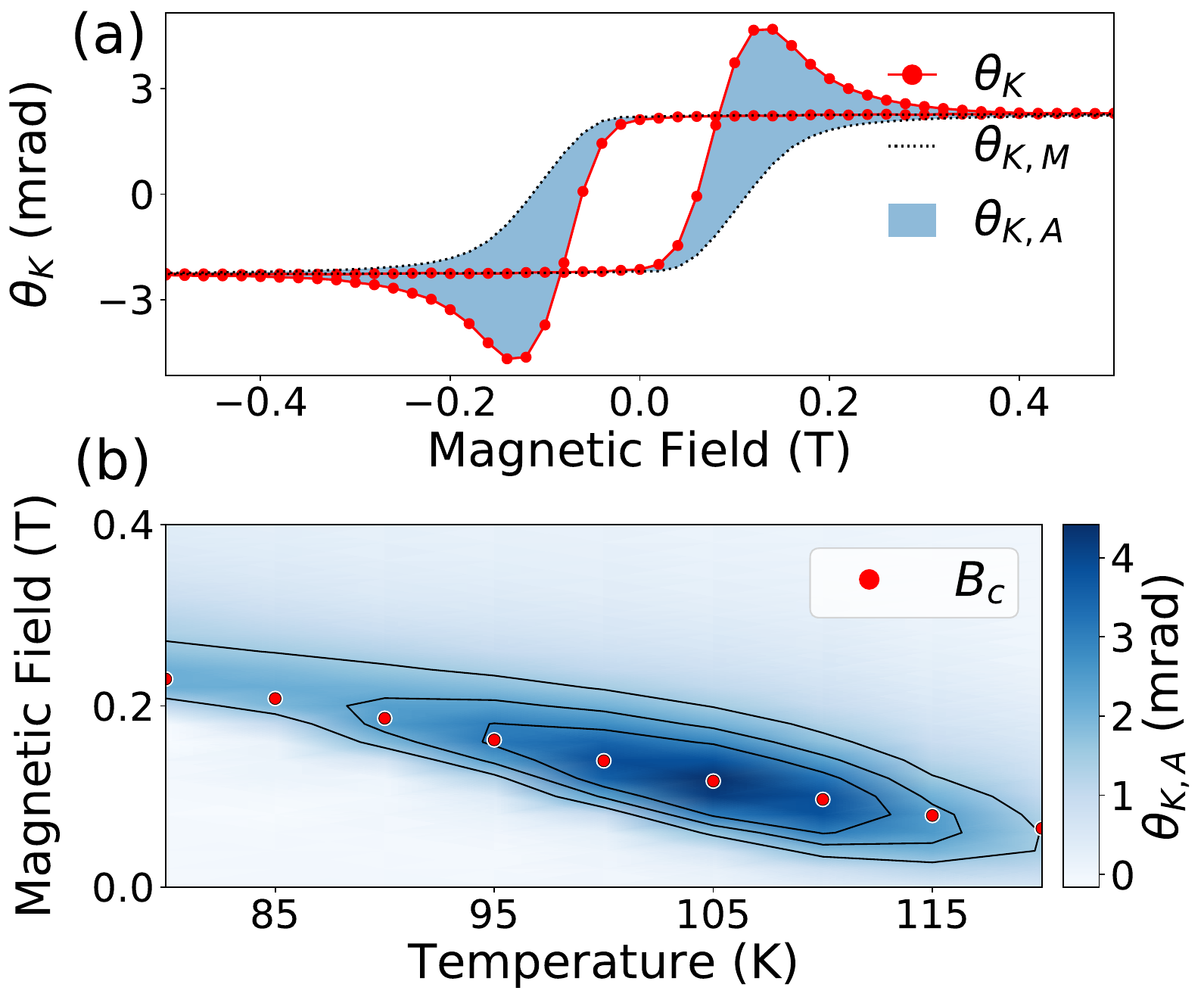}
 \caption{ (a) Kerr rotation measured with a $600$\,nm laser at $T\!=\!105$\,K for an $88$\,nm SrRuO$_3$
 film on LSAT, showing the expected signal for a linear dependence on magnetization $\theta_{K,M}$, and the additional contribution $\theta_{K,A}$. (b) Colour plot of the 
 $\theta_{K,A}$ in the temperature-field plane. Red dots indicate $B_C$, the field strength where the net magnetization becomes zero. \label{fg:phase}}
\end{figure}

For the MOKE data in which anomalies are present, such as those in Figs.~1(c) and 1(d), the Kerr angle $\theta_K$ appears to exhibit two components: one component tracks the 
$M$ hysteresis loop (with the proportionality constant potentially changing with temperature, as discussed above), while the second component produces the bump-like anomaly. To see this more clearly, we can separate $\theta_K$ 
into a corresponding ``normal contribution'' $\theta_{K,M}$, and an additional contribution $\theta_{K,A}$, which we call the ``anomalous Kerr angle'', 
so that $\theta_K = \theta_{K,M} + \theta_{K,A}$.
We illustrate this analysis in Fig.~\ref{fg:phase} for $\lambda\!=\!600$\,nm at $T=105$\,K, where we obtain $\theta_{K,M}$ by scaling directly measured magnetization data in order to cancel off the Kerr rotation at saturation magnetization [see dashed line in Fig.~\ref{fg:phase}~(a)]. Subtracting this from the $\lambda\!=\!600$\,nm Kerr signal loop (line with dots) gives a difference signal $\theta_K - \theta_{K,M}$ (blue shaded area) which defines the anomalous Kerr angle $\theta_{K,A}$. 
The full field and temperature dependence of $\theta_{K,A}$ quantified in this manner 
is displayed in Fig.~\ref{fg:phase}(b); the dots indicate the coercive field $B_c$ where the net magnetization vanishes. Here we only show temperatures above 80 K; $\theta_{K,A}$ appears to persist down to very low temperatures but is difficult to clearly extract in this manner, as it becomes relatively small while $\theta_{K,M}$ becomes much larger.
We observe that the largest $\theta_{K,A}$ occurs roughly around $B_c$. It is tempting to assign this feature to skyrmions as they form
in the vicinity of magnetization reversals \cite{Panagopoulos_NComm2019, Meng2019} in ultrathin films with chiral magnetic interactions.
However, SrRuO$_3$ films in our thickness regime possess no known mechanism to stabilize skyrmions, making this an implausible explanation.

\begin{figure}[t]
 \includegraphics[width=\columnwidth]{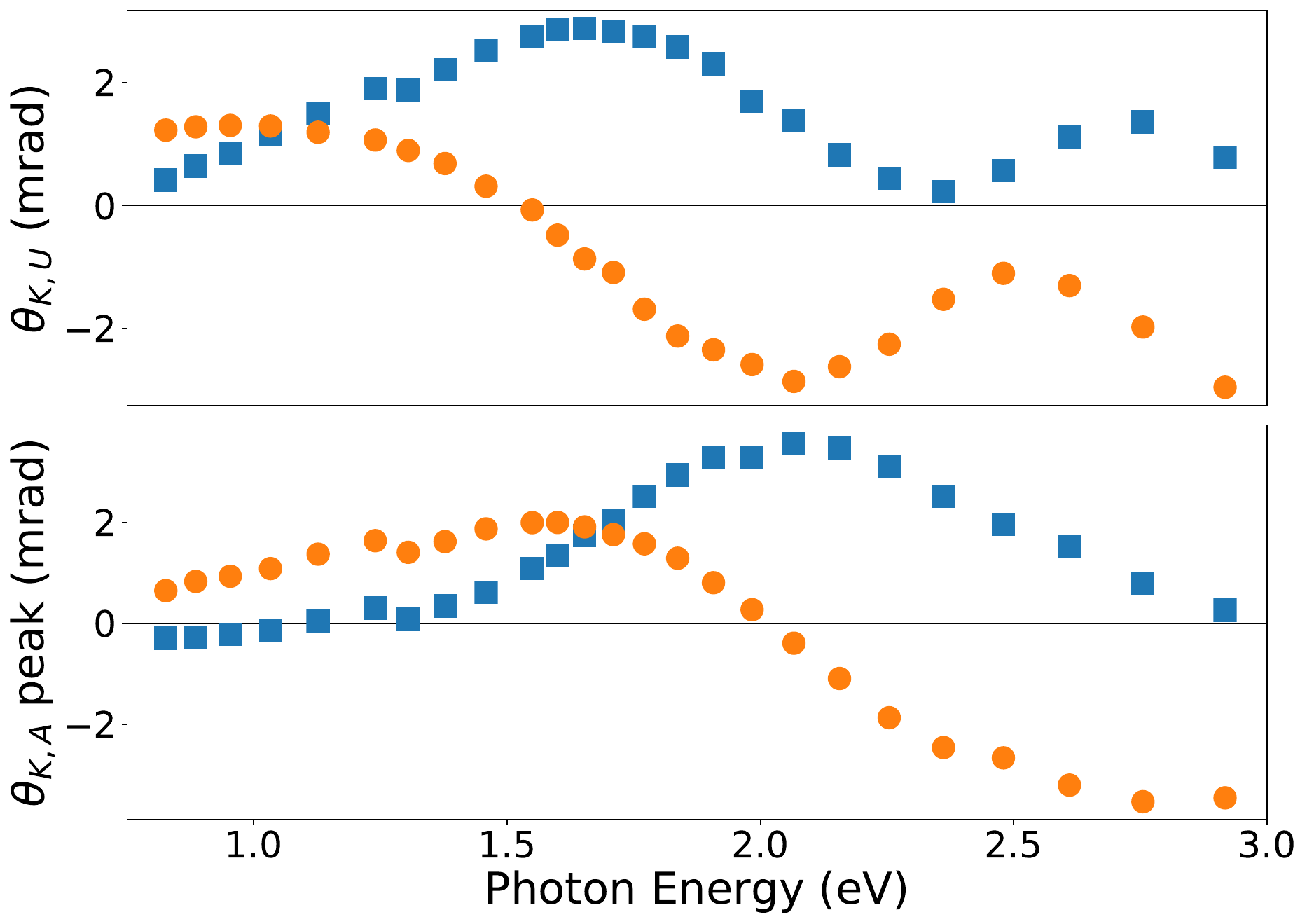}
 \caption{Real (blue squares) and imaginary (orange circles) parts of the Kerr rotation at uniform magnetization $\theta_{K,U}$ (top) and the peak of the anomalous Kerr angle $\theta_{K,A}$ (bottom), shown as a function of frequency, for $T=110$ K.\label{fg:wavelengths}}
\end{figure}

We clearly observe the frequency dependence of these anomalies with this analysis. Figure~\ref{fg:wavelengths} shows both the Kerr rotation at uniform magnetization $\theta_{K,U}$ (as discussed earlier) and the peak value of the anomalous component, $\theta_{K,A \text{ peak}}$, as a function of the photon energy of the laser. Here we also show the imaginary part of the Kerr angle (see SM for more details). We note that the appearance of the anomalies exhibits a resonant behaviour, with a peak at around $\lambda\!=\!600$\,nm (2 eV) in the real part, and a concomitant zero crossing in the imaginary part consistent with Kramers-Kronig relations. 

Similar measurements on films of varying thicknesses yield a nonzero $\theta_{K,A}$ for thicknesses in the range $\sim 70$-$120$\,nm; see 
SM \cite{suppmat} for additional data. In some thinner films ($\sim\! 30$\,nm), we observe anomalies in 
$R_{xy}$, but not in any of our MOKE measurements. However, in these cases we also do not observe any zero crossings of $\theta_{K,U}$ 
at non-zero magnetization for the measured $\lambda$, as for the long wavelength data in the $88$\,nm film. 
(It is possible that the anomalies may appear at $\lambda$ outside the range accessed in our study.)
Finally, films grown on LSAT which has greater lattice mismatch (1.4\%) 
with SrRuO$_3$ when compared  with the STO substrate (lattice mismatch: 0.45\%), exhibit a larger, and broader (in field), $\theta_{K,A}$. Thus
strain has a strong impact on the observed anomalies. 

In previous work on $R_{xy}$ anomalies in ultrathin SrRuO$_3$ films, it was suggested that an alternative source (other than skyrmions) for the bumps
could be the temperature-dependent sign changes in the bulk anomalous Hall effect combined with inhomogeneities in $T_c$ or the coercive field
across the sample \cite{AlternateExplanation}. 
While such inhomogeneities could be important in few unit-cell ultrathin films, where they were proposed to originate from single unit-cell
variations in the film thickness, it is less clear that such atomic scale variations can impact the relatively
thick films studied in our work. However, a ubiquitous feature common to such magnetic thin films is
magnetic domain proliferation near the coercive field \cite{Meng2019, MFM2}. 
We thus turn to a phenomenological theory of the MOKE in the presence of such domains.

{\it Theory. ---}
As we pass through the coercive field $B_c$ during a magnetization reversal process, minority magnetic domains start to proliferate, and eventually take over the
system. During such a process, let ${\cal M}(B,\br)$ be the inhomogeneous local magnetization (perpendicular to the film) at a point  $\br$ in a field $B$. The average 
magnetization $M(B) \!=\! (1/V) \! \int \!d^3 \br~ {\cal M}(B,\br)$, where $V$ denotes the sample volume. Let
the Kerr angle in a system with {\it uniform} magnetization $M$ be given by $\theta_{K,U} (M)$. We then propose that the measured Kerr angle
$\theta_K (B) \!=\! (1/V)\! \int \! d^3\br ~\theta_{K,U}({\cal M}(B, \br))$; this averaging result in a nontrivial behavior when $\theta_{K,U}(M)$ is a nonlinear
function.
A heuristic justification for this local magnetization approximation (LMA) for
$\theta_K$ is that the electronic response at high frequency $\omega$ must be local in space. Semiclassically, electrons with Fermi velocity $v_F$
traverse only a distance $\ell_\omega \! \sim \! \pi v_F/\omega$ in a half-period $\pi/\omega$; 
using $v_F \!\sim\! 2\!\times\! 10^5$\,m/s \cite{Fermi_velocity}
and $\hbar\omega \!\sim\!2$\,eV, 
yields $\ell_\omega\! \sim\! 2$\,\AA, i.e., on the scale of the lattice spacing. In the SM \cite{suppmat}, we
compare the LMA with the numerically computed frequency-dependent conductivity tensor from the Kubo formula for a $t_{2g}$ model with spin-orbit coupling 
using various inhomogeneous magnetization profiles, showing that they quantitatively agree for $\hbar\omega \!\gtrsim\! W/4$, 
where the $t_{2g}$ bandwidth $W \!\sim\! 2$-$3$\,eV \cite{Fang2003}.

\begin{figure}[b]
\centering
\includegraphics[width=\columnwidth]{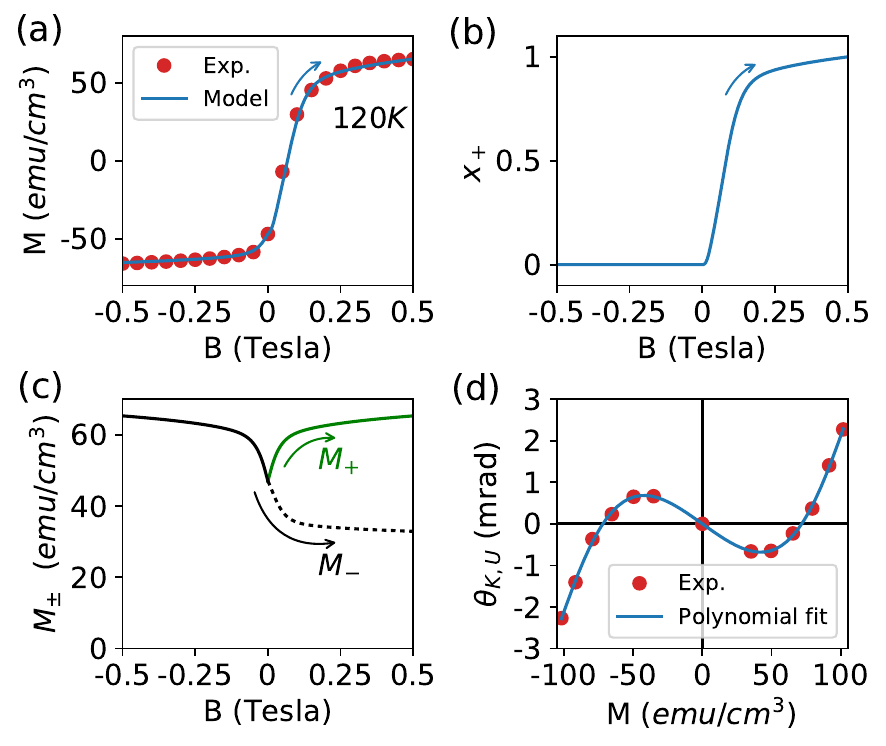}
\caption{
(a) Magnetization data (dots) at $T\!=\!120$\,K in an increasing-field sweep for $88$\,nm SRO film.
Solid line is a fit to the two-domain model $M \!=\! x_{+} M_+ \!-\! x_{-} M_-$ discussed in the text and SM \cite{suppmat}.
Extracted domain fraction $x_+$, and domain magnetizations $M_{\pm}$ are shown in (b) and (c), with arrows indicating 
the direction of the field sweep. (d) Kerr angle $\theta_{K,U}$ for $\lambda\!=\! 600$\,nm as a function of uniform magnetization
for $88$\,nm film; 
dots are data points are obtained in a large applied field at 
different temperatures, solid line shows a fit.}
\label{fig:theory1}
\end{figure}

We can then compute the Kerr loop in three steps. (i) We simplify the inhomogeneous magnetization profile ${\cal M}(B,\br)$ by two domain types,
with magnetizations $M_{+}(B)$ and $-M_{-}(B)$ perpendicular to the film, and corresponding volume fractions
$x_{+}(B)$ and $x_{-}(B)\!=\!1\!-\!x_{+}(B)$. Here $M_{\pm}\geqslant 0$ and $0\leqslant x_{\pm}\leqslant 1$. This yields $M(B) \!=\! x_+ (B) M_+ (B) - x_{-}(B) M_-(B)$. 
We fit the magnetization data to extract $x_+,M_+,M_-$ as a function of $B$.
(ii) Next, we assume that the Kerr angle at saturation magnetization is a good proxy for $\theta_{K,U}$. 
We thus fit the saturation Kerr angle as a function of saturation magnetization to get $\theta_{K,U}(M)$.
(iii) Finally, using these fitted functions as inputs to the LMA, we obtain the theoretical estimate for the average
Kerr angle as $\theta_K(B) \!=\! x_+(B) \theta_{K,U}(M_+(B)) + x_{-}(B) \theta_{K,U}(-M_-(B))$.

\begin{figure}
\centering
\includegraphics[width=\columnwidth]{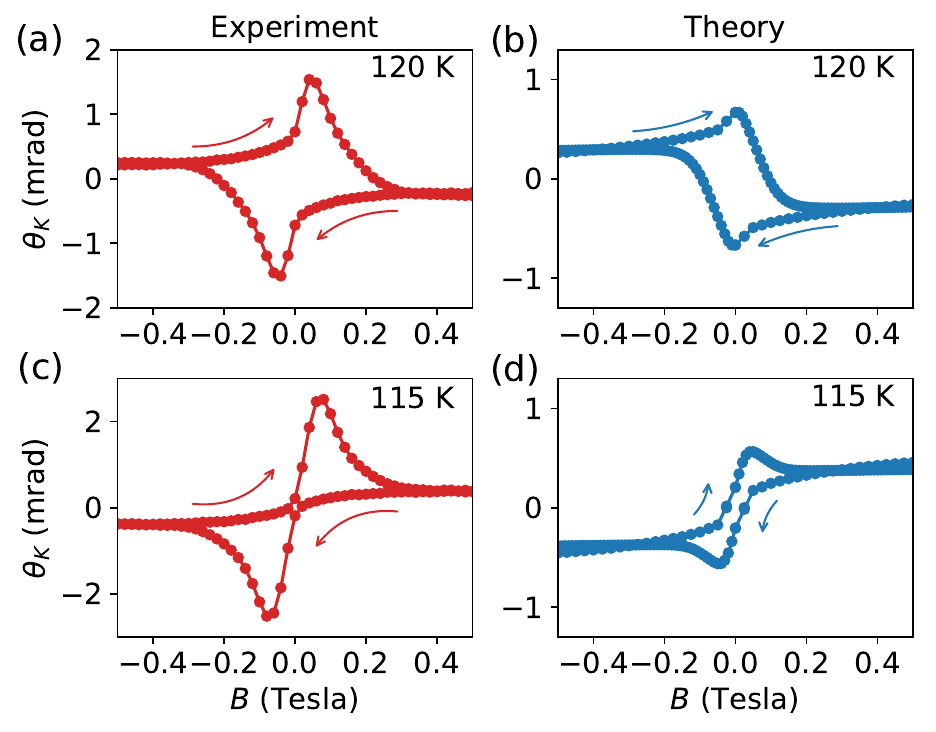}
\caption{(a) and (c) depict Kerr anomalies seen in experiments on the 88 nm SrRuO$_3$ sample using a 600 nm laser at 120 K and 115 K 
respectively. Arrows denote the direction of the field sweep. 
(b) and (d) show the theoretically computed Kerr anomalies within the LMA, displaying bump features which agree qualitatively with the 
experiment. The anomalies arise from magnetic domains and the nonmonotonic magnetization dependence of $\theta_{K,U}$ fits shown in Fig.~\ref{fig:theory1}(d).}
\label{fig:theory2}
\end{figure}

Figure~\ref{fig:theory1} (a) shows an example of a fit to the magnetization data in an increasing-field sweep (see SM \cite{suppmat}
for details). We make a simple ansatz for $x_+$, with $x_+(B \!<\! 0) \!=\!0$, and $x_+\!\to \!1$ for large $B \!>\!0$. Thus, 
$M_-(B \!<\! 0) = -M$ which can be read off from the measured $M$, shown as the solid black line in 
Fig.~\ref{fig:theory1}(c). Symmetry dictates $M_+(B\! >\! 0)\!=\! M_-(B \!<\! 0)$, as shown by the
solid green line. These two constraints, along with a reasonable ansatz for $M_-(B \!>\! 0)$, allow us to fit $M(B\!>\!0)$, 
and extract $x_+(B \! > \!0)$ 
and $M_-(B \!>\!0)$ as shown in Fig.~\ref{fig:theory1}(b)-(c). 
Our results below are robust against variations in 
the precise shape of $M_-(B \!>\! 0)$.

We next turn to the saturation Kerr angle, $\theta_{K,U}(M)$, which is measured at a large enough field $|B| \!\sim\! 0.5$\,T
to achieve saturation at a given temperature,
and then varying the temperature to tune the  saturation value of $M$.  Figure~\ref{fig:theory1}(d) shows a polynomial fit to $\theta_{K,U}(M)$,
which exhibits a non-monotonic behavior similar to the Weyl-node induced 
non-monotonic Hall resistivity \cite{Fang2003, Nagaosa2010,Burkov2013}.

These key results from the fits in Fig.~\ref{fig:theory1} serve as inputs to the LMA. Figure~\ref{fig:theory2} compares the experimentally measured $\theta_K$
and the LMA theory results over the hysteresis loop, showing
good qualitative agreement in the overall shape and magnitude of the observed bumps.
Achieving a better quantitative agreement in terms of the height of 
the bumps and the precise shapes requires a theory for $\theta_{K,U}(M)$ at each temperature, while we have
extracted it using experimental values from different temperature datasets; see SM \cite{suppmat} for further discussion of this point.

Such an effective medium approximation fails to explain the anomalies we observe in $R_{xy}$ \cite{suppmat};
this discrepancy may be due to the fact that the d.c. conductivity tensor is non-local and expected to be more sensitive to details of the domain size 
distribution and domain walls. Using a $t_{2g}$ model Hamiltonian, we show in the SM \cite{suppmat}
that domain walls introduce a correction to $\rho_{xy}$ which can potentially
account for the Hall anomalies.

{\it Conclusion. ---} We have measured an anomalous contribution to the Kerr signal in thin films of SrRuO$_3$ over wide regimes of magnetic field,
temperature, and laser wavelength, which appears 
to behave similar to $R_{xy}$ anomalies attributed to the THE in skyrmion materials.
We have instead shown that these anomalies in the Kerr signal can arise from the non-linear dependence of the Kerr angle on magnetization,
together with magnetic domain formation during magnetization reversal.
For bulk Hall transport, previous work has shown that the nonmonotonic behavior of the Hall resistivity as a function of magnetization or temperature
can arise from Weyl nodes in the band structure \cite{Nagaosa2010,Burkov2013}. Weyl nodes, and their interplay with magnetic domain walls, 
might thus provide a microscopic origin for the non-linear Kerr signal as well as the observed d.c. Hall anomalies. Such an interplay 
has been recently examined for $R_{xy}$ in the antiferromagnetic Weyl metals Mn$_3$Ge and Mn$_3$Sn \cite{BalentsPRL2017Mn3X,BehniaNComm2019Mn3Sn}.  
In light 
of our results, it would be 
valuable to also reevaluate the role of magnetic domains versus skyrmions in ultrathin SrRuO$_3$ films.

\begin{acknowledgments}
The optical measurements were performed at Tsinghua University and at the University of Toronto and were supported by the Tsinghua University Startup Fund, the CIFAR Azrieli Global Scholars Programme, NSERC Canada Research Chair, the Canadian Foundation for Innovation, and the Ontario Research Fund. The theoretical studies were performed at the University of Toronto and were funded by NSERC of Canada. This research was enabled in part by support provided by WestGrid (www.westgrid.ca) and Compute Canada Calcul Canada (www.computecanada.ca). Sample growth and characterization were carried out at Tsinghua University and were supported by the Basic Science Center Project of NFSC under grant No. 51788104; the National Basic Research Program of China (grants 2015CB921700 and 2016YFA0301004); and the Beijing Advanced Innovation Center for Future Chip (ICFC).
\end{acknowledgments}

\clearpage

Supplemental Material

\textbf{TABLE OF CONTENTS}\\
\textbf{S1. Additional MOKE data for the 88 nm sample at different laser frequencies}\\
\textbf{S2. MOKE data on different thickness films on STO and LSAT substrates}\\
\textbf{S3. MOKE measurements at oblique incidence}\\
\textbf{S4. Alternative model for Hall measurements}\\
\textbf{S5. Local magnetization approximation (LMA) for optical response}\\
\textbf{S6. Ansatz for the two-domain model}\\
\textbf{S7. Impact of modifying the shape of the $\theta_{K,U}$ curves}\\
\textbf{S8. Impact of domain walls on dc Hall conductivity}\\


\section{S1. A\lowercase{dditional} MOKE \lowercase{data for the 88 nm sample at different laser frequencies}}

For the 88 nm SrRuO$_3$ on LSAT, full datasets of both the real ($\theta_K$) and imaginary ($\epsilon_K$) parts of the Kerr angle as a function of field and temperature were taken at five different wavelengths - 500 nm, 600 nm, 700 nm, 800 nm, and 1500 nm. Here, we show a series of plots analogous to the one shown in Figure 2 of the main text, where an example of raw data at $T=105$ K is shown along with a color plot showing the anomalous part of the signal as a function of both field and temperature, for each of these wavelengths, as well as for the Hall resistivity (Fig. S1).

\begin{figure}[h]
	\includegraphics[width=0.9\columnwidth]{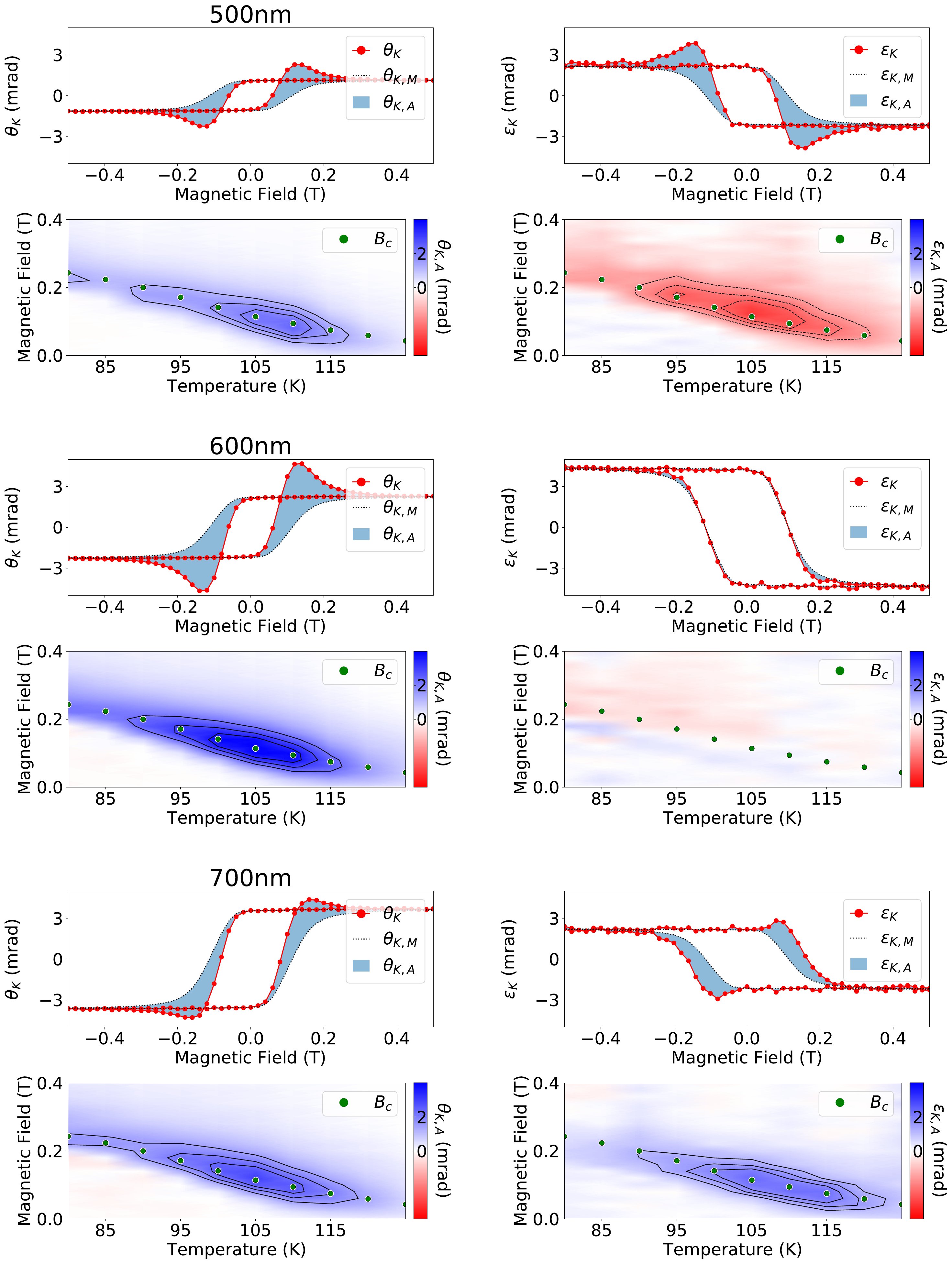}
\end{figure}
\clearpage
\begin{figure}[h]
	\includegraphics[width=0.9\columnwidth]{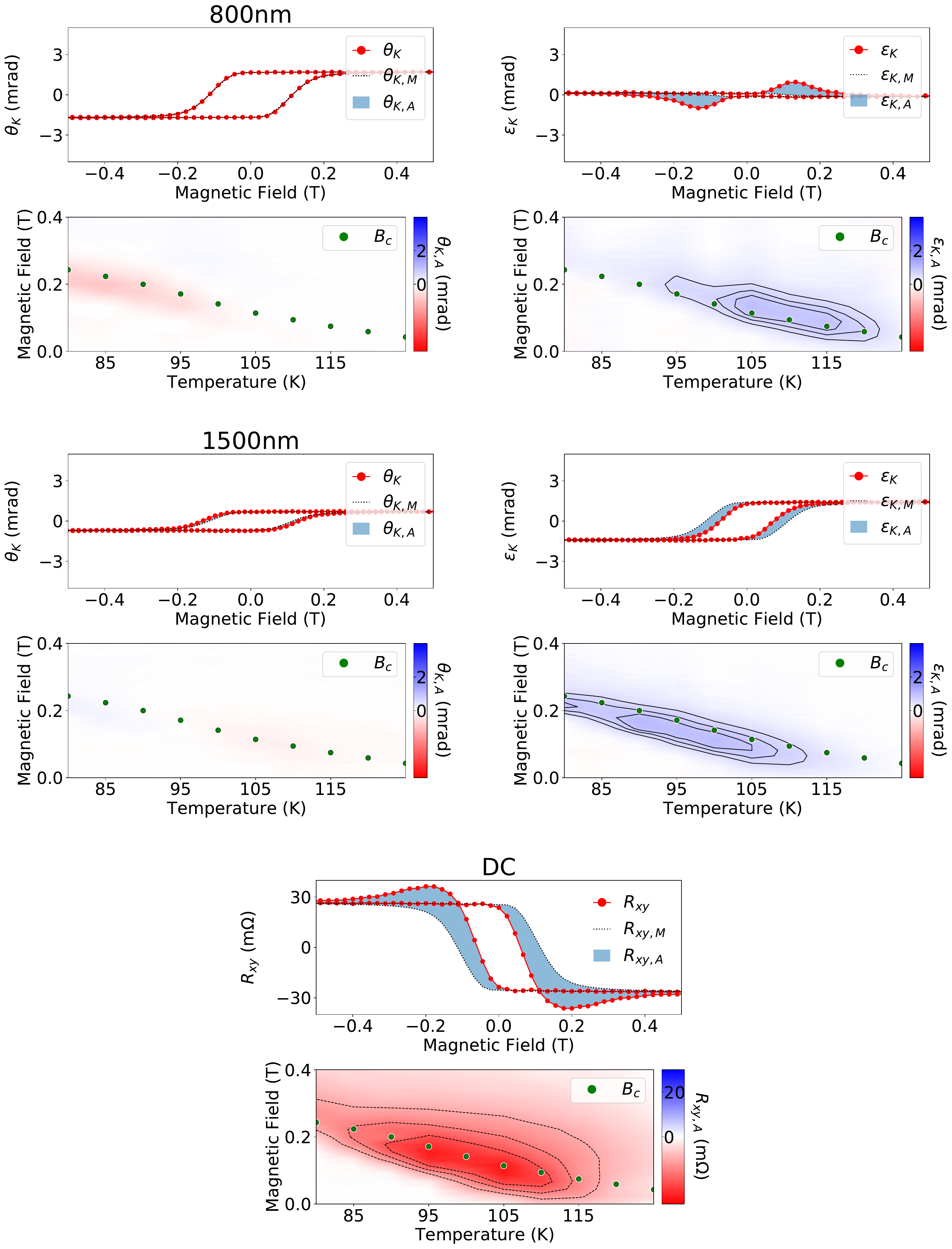}
	\caption{Plots showing the presence of anomalies in the Kerr rotation (real part: $\theta_K$, imaginary part: $\epsilon_K$) for various wavelengths and Hall resistivity ($R_{xy}$). The top plot of each set shows a hysteresis loop at $T=105$ K and its division into the normal and anomalous components, and the bottom shows the anomalous component as a function of field and temperature. \label{colourplots} }
\end{figure}
\clearpage

\section{S2. MOKE \lowercase{data on different thickness films on} STO \lowercase{and} LSAT \lowercase{substrates}}

\begin{figure}[h]
 \includegraphics[width=\columnwidth]{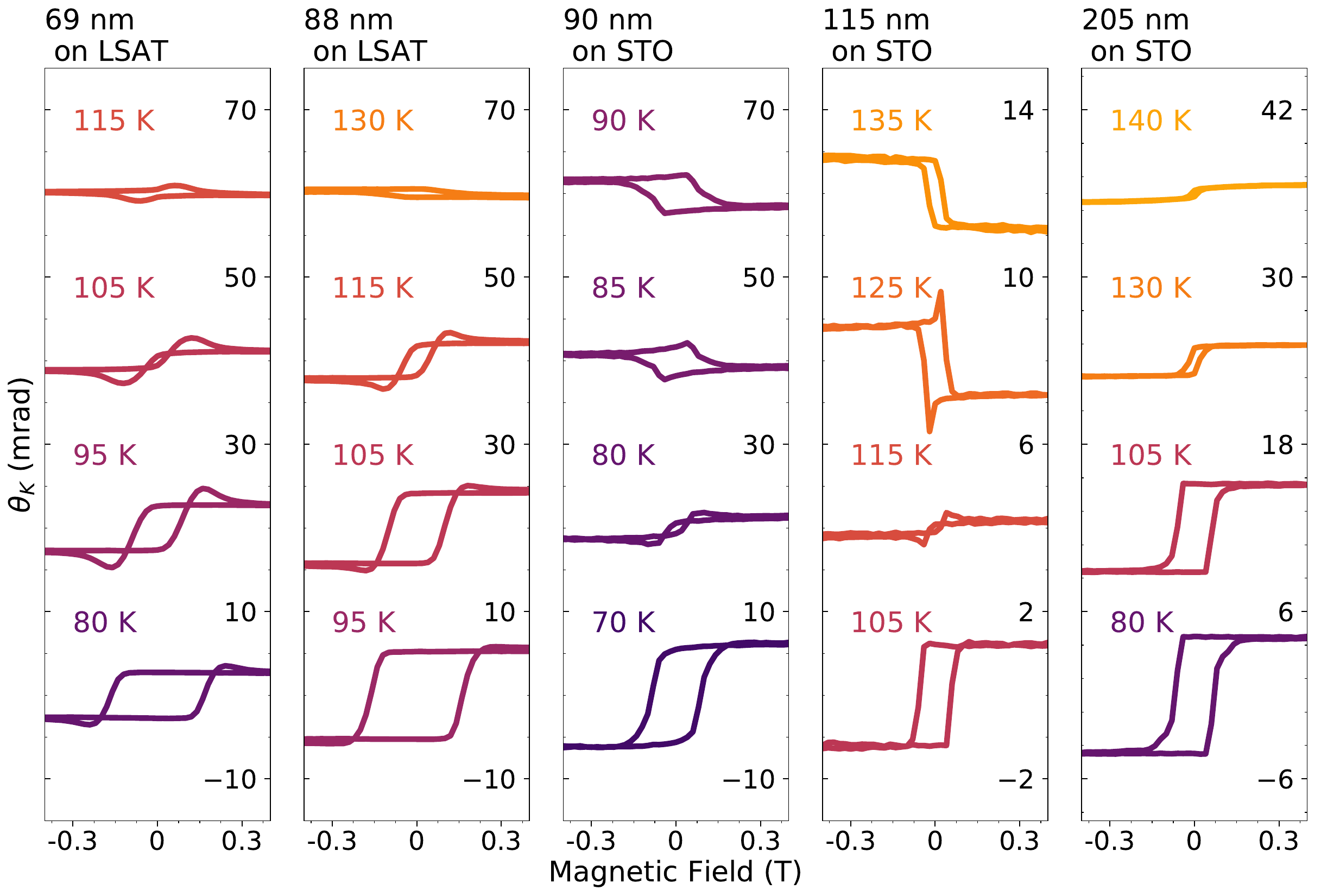}
 \caption{Data at selected temperatures for various samples using a HeNe laser at 633 nm. A notable difference in the samples on STO is that they have much smaller and narrower bumps compared to those on LSAT. \label{fg:samples}}
 \end{figure}

The MOKE signal was strongly affected by the thickness of the samples, as well as the choice of substrate. Figure \ref{fg:samples} shows data for a few different samples. Between the two substrates, the STO samples show much smaller and narrower bump features. Between the different thicknesses shown, the difference was mostly in what the temperature range the anomalies occurred, although going to thinner samples (the next being a 51 nm thick sample on LSAT) resulted in these features disappearing entirely. Note that this data set only includes the real part of the Kerr angle, as it was taken on a different setup where only the real part was measured, rather than the full complex Kerr angle as in the other data sets.

\section{S3. MOKE \lowercase{measurements at oblique incidence}}

\begin{figure}[h]
 \includegraphics[width=0.9\columnwidth]{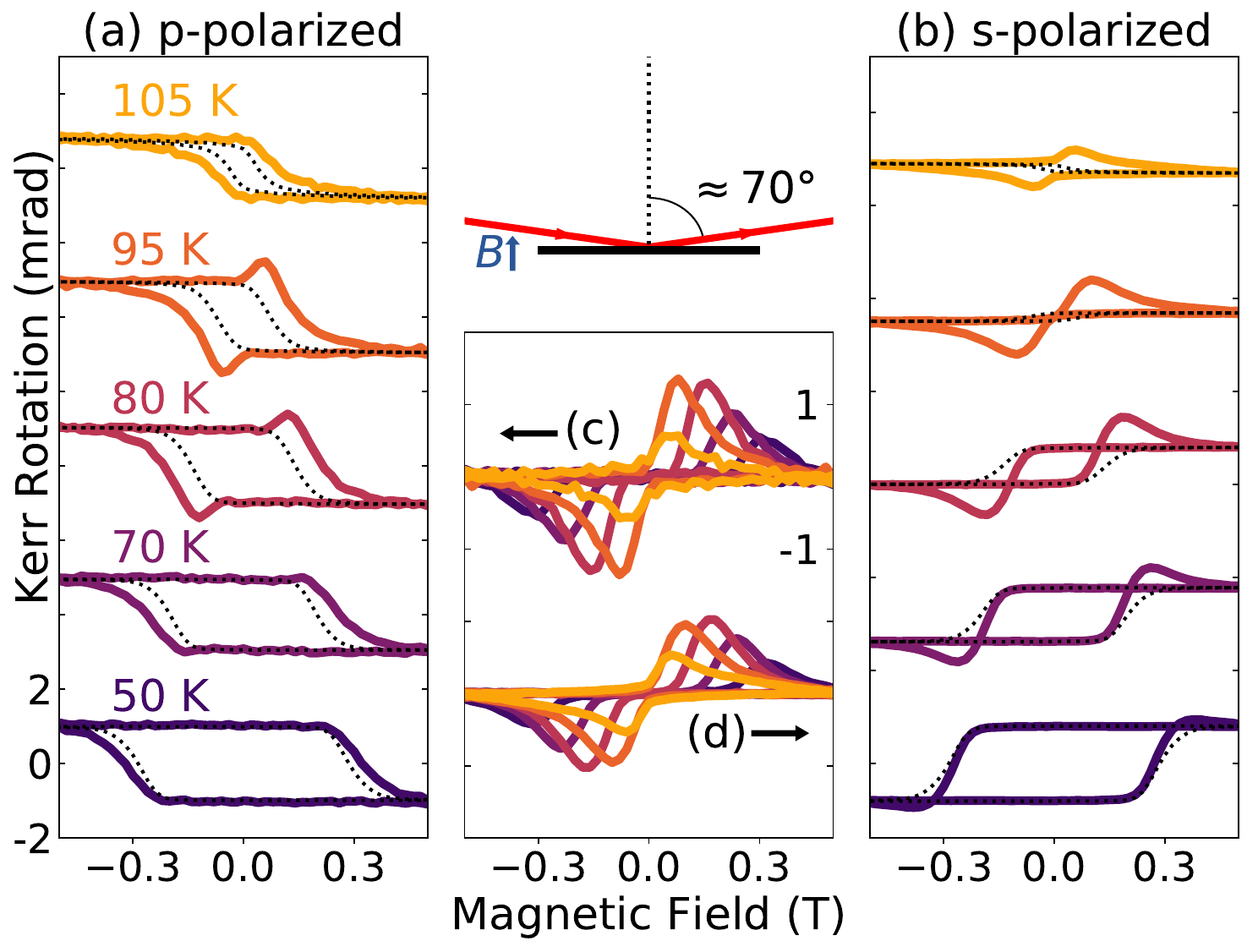}
 \caption{The real part of MOKE signals taken with an oblique angle of incidence using a 633 nm laser, for a 120 nm SrRuO$_3$ sample on LSAT. (a) and (b) show the full results with p- and s-polarization, respectively, and (c) and (d) show the anomalous parts $\theta_{K,A}$ after subtraction of $\theta_{K,M}$. \label{fg:oblique}}
 \end{figure}

In addition to data taken in the standard polar Kerr configuration, we acquired some data where the laser was reflected from the sample at a large angle of incidence of around 70 degrees while the magnetic field was applied perpendicular to the film as shown schematically in Fig. \ref{fg:oblique}. The s-polarized data (b) is similar to that shown previously, with $\theta_{K,M}$ changing sign just above 95 K. 
The p-polarized data (a), on the other hand, has $\theta_{K,M}$
which stays fairly constant throughout the temperature range shown. In both cases the additional contribution $\theta_{K,A}$ 
is clearly visible, as shown in (c) and (d). Despite the dramatic differences in the temperature dependence of $\theta_{K,M}$, the additional 
contribution $\theta_{K,A}$ remains similar in sign and magnitude between the different polarizations.  
Because in our model the anomalies only come from nonlinearity in the $\theta_K$ versus $M$ relationship, this does not necessarily pose a problem. However, we do note that in the usual expressions for a simple out-of-plane magnetization, the p- and s-polarized data would be related by a single overall factor (determined by the angle of incidence and index of refraction), which does not appear to be the case here. This could be due to the presence of in-plane magnetization, as the magnetization axis has been reported to tilt to roughly $30-45$ degrees out of plane  \cite{rmpsro}.

\clearpage
 
 \section{S4. A\lowercase{lternative model for} H\lowercase{all measurements}}
 Recently it was suggested that the Hall effect data on ultrathin SrRuO$_3$ films could be explained by inhomogeneity characterized by a distribution of effective temperatures \cite{AlternateExplanation}. Here we discuss why such a ``non-intrinsic" explanation of the data in terms of inhomogeneities does not work. 
 
\begin{figure}[h]
 \includegraphics[width=\columnwidth]{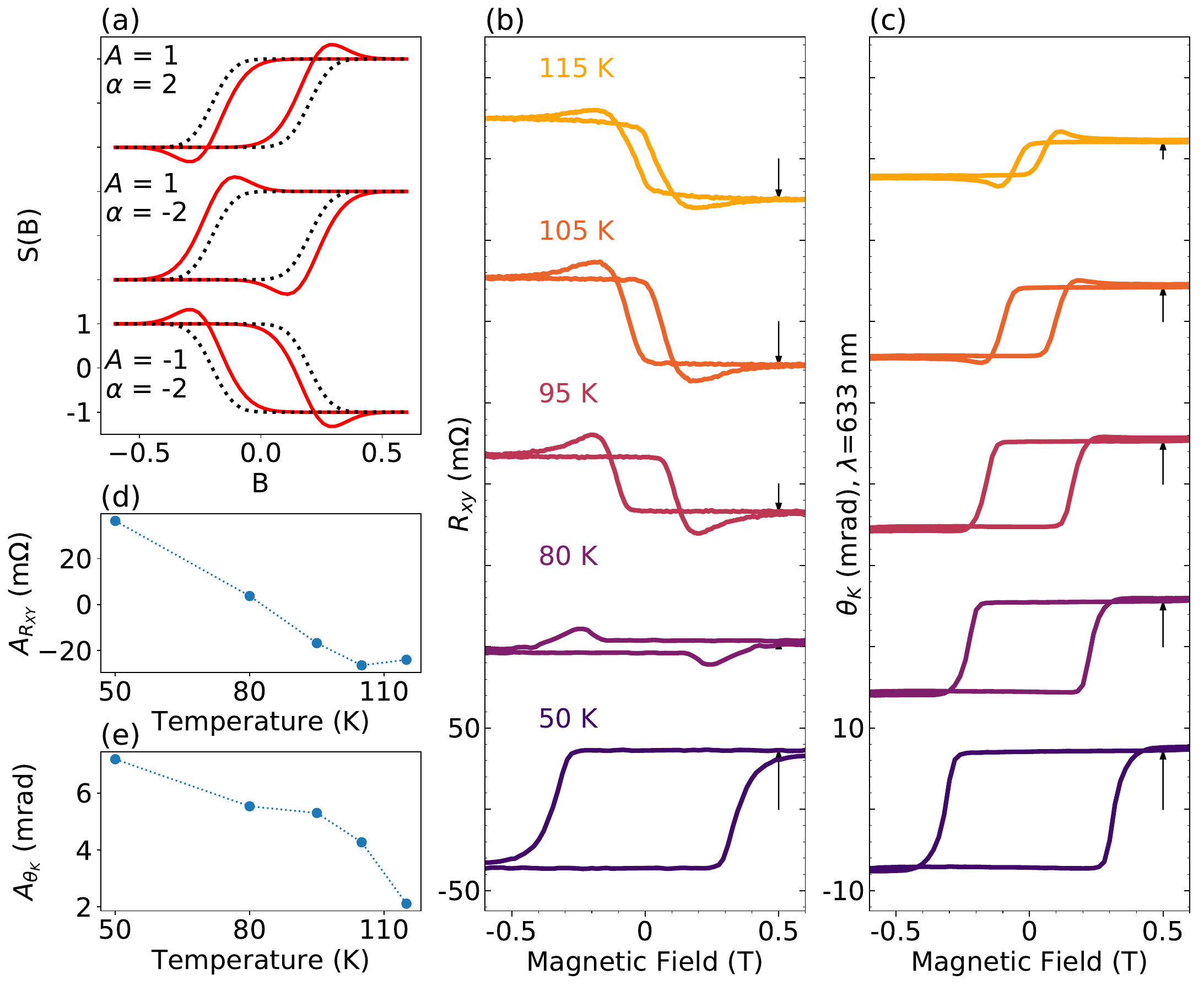}
 \caption{(a) Example plots generated using Equation \ref{eq:alt} and various $A$ and $\alpha$ values, with $B_c = 0.2$ and $\beta=0.1$ being held constant. (b) Hall and (c) MOKE data for an 88 nm thick sample on LSAT. Black arrows indicate the saturation values, which are plotted in (d) and (e) as a function of temperature. In both cases, the saturation values are clearly decreasing with temperature, which means $\alpha < 0$. In the $R_{XY}$ data, the bumps have a negative amplitude, matching the prediction of Equation \ref{eq:alt}, but in $\theta_K$ they are positive, making the data impossible to reproduce with this model. \label{fg:alternativemodel}}
 \end{figure}

 As a simple model, we can write the signal under increasing field as a step function with a temperature dependent saturation value $A(T)$ and coercive field strength $B_c(T)$
 \begin{equation}
 	S_0(B,T) = A(T) \Theta(B - B_c(T)).
 \end{equation}
We then take the observed signal to be a combination of signals at different temperatures, characterized by a normal distribution with a width $\sigma_T$
\begin{equation}
	S(B,T) = \int \text{norm}(\widetilde{T}/\sigma_T)S_0(B,T+ \widetilde{T})d\widetilde{T}.
\end{equation}
Taking the temperature dependence of $A$ and $B_c$ to be approximately linear, and defining parameters $\alpha$ and $\beta$ so that $A(T+\widetilde{T}) \approx A(T) + \alpha(T) \widetilde{T} / \sigma_T$ and $B_c (T+\widetilde{T}) \approx B_{c}(T) - \beta(T) \widetilde{T} / \sigma_T$, where $\beta$ is taken to be positive (as is the case physically), gives a solution
 \begin{equation}
 	S(B,T) = A(T) \cdot \text{erf}\Big(\frac{B-B_{c}(T)}{\beta(T)}\Big)  + \alpha(T) \cdot \text{norm}\Big(\frac{B-B_{c}(T)}{\beta(T)}\Big), \label{eq:alt}
 \end{equation}
 where erf and norm are the standard error function and normal distribution respectively. The first term produces a typical magnetization loop, and the second produces features that closely resemble the observed bumps. The constraint of this model, however, that the amplitude and sign of these bump features are set by $\alpha$, as shown in Fig. \ref{fg:alternativemodel} (a). In our data [Fig. \ref{fg:alternativemodel} (b)-(e)] the saturation value of the signal decreases with temperature for both $R_{XY}$ and $\theta_K$, which means that $\alpha$ must be a negative number. While the bumps in the $R_{XY}$ data do indeed also have a negative amplitude, those seen in $\theta_K$ have a positive amplitude and therefore cannot be explained by the model presented above.

\clearpage 

\section{S5. L\lowercase{ocal magnetization approximation} (LMA) \lowercase{for optical response}}
\label{appdx_lma}
The LMA as used in the main text for the measured Kerr angle states that
\bea
\theta_K^{LMA} &=& \frac{1}{V}\int d\br ~\theta_{K, U}(\mathcal{M}(\br)),
\label{eq:LMA_kerr}
\eea
where $\theta_{K, U}(M)$ is the Kerr angle for a fixed spatially uniform magnetization $M$. This approximation
relies on the observation that the conductivity tensor at high frequencies (i.e., optical frequencies) is expected to be
spatially local, and may thus be obtained by averaging the local conductivity tensor across the system. In this section, we provide numerical evidence in support of this
approximation, by comparing the conductivity tensor for a system with inhomogeneous magnetization calculated (i) using the LMA, i.e., $\sigma_{ij}^{LMA}$, with (ii)
the exact result computed directly using the Kubo formula.

We consider a cubic-lattice $t_{2g}$ electron Hamiltonian, which has been used to study the anomalous Hall effect in metallic ferromagnets \cite{Burkov2013},
and introduce a spatially varying Weiss field $w(i)$ which produces an inhomogeneous exchange splitting and magnetization:
\bea
H &=& H_{\text{hop}} + \lambda_{\rm soc} \sum_i \mathbf{\hat{L}}_i \cdot \mathbf{\hat{S}}_i - \sum_i w(i) S_{z, i},\\
\label{eq:t2g_hamiltonian}
H_{\text{hop}} &=& - \sum_{\langle ij \rangle} t_{ij, a} d^{\dagger}_{ia\sigma}d_{ja\sigma} + \text{h.c.} - \sum_{\langle \langle ij \rangle\rangle} f_{ij, ab} ~d^{\dagger}_{ia\sigma}d_{jb\sigma} + \text{h.c.},
\eea
where $\lambda_{\rm soc}$ is the on-site spin-orbit coupling strength, $\boldsymbol{\hat{L}}$ is the angular momentum operator in the $t_{2g}$ basis, and $\mathbf{\hat{S}}$ is the spin operator. 
In $H_{\rm hop}$, the first term $t_{ij}$ denotes the amplitudes for intra-orbital nearest-neighbor hopping, while $f_{ij}$ refers to next-neighbor inter-orbital hopping amplitudes, 
and we sum over the orbital index $a = 0, 1, 2$ which corresponds respectively to the $yz, zx$ and $xy$ orbitals. 

The hopping integral $t_{ij, a}$ is labelled by the bond index $ij$, which can take three values corresponding to the unit vectors in the x, y and z-direction, i.e., $\mathbf{\hat{e}}_b$ for $b = 0, 1, 2$. Explicitly, $t_{ij, a} \equiv t_{\hat{e}_b, a} = (1-\delta_{ab})t_1 + \delta_{ab} t_2$. The inter-orbital hopping occurs
on six bonds: $\mathbf{\hat{e}}_1 \pm \mathbf{\hat{e}}_2$, $\mathbf{\hat{e}}_2 \pm \mathbf{\hat{e}}_3$ and $\mathbf{\hat{e}}_3 \pm \mathbf{\hat{e}}_1$. The corresponding
hopping integrals are given by $f_{ij, ab} = f_{\hat{e}_c \pm \hat{e}_d, ab} = \pm f ~ (\delta_{ac}\delta_{bd}  + \delta_{ad}\delta_{bc})$. In the rest of this section, 
we discuss the results for the case where the Weiss field $w(i)$ is chosen to be uniform in the $\hat{z}$ direction, but has strong inhomogeneity in the $xy$-plane. We do this
by going to momentum space assuming a super-cell having unit-cell dimension $N_x \times N_y \times 1$.

The Kubo formula for the conductivity tensor $\sigma_{ij}(\omega)$ of the inhomogeneous (super-cell) system  is given by \cite{mahan2000, *coleman,*zhang}:
\bea
\sigma_{ij} &=& \frac{\I 2\pi}{V}\frac{e^2}{h} \sum_{\bk, l, m} \frac{n_F(E_{\bk m}) - n_F(E_{\bk l})}{E_{\bk l} - E_{\bk m}} \left[\frac{(v_i)_{ml}(v_j)_{lm}}{\hbar\omega+i\gamma+(E_{\bk m} - E_{\bk l})}\right],
\label{eq:kubo_formula}
\eea
where $(v_i)_{ml} \equiv \bra{\bk m}\frac{\partial \mathcal{H}(\bk)}{\partial k_i}\ket{\bk l}$ are matrix elements of the velocity operator, 
and $\ket{\bk m}$ is a Bloch state with an eigenvalue $E_{\bk m}$ given the super-cell configuration,
$n_F$ is the Fermi function and $\gamma$ is a small broadening. 

We will compare this with the LMA response tensor, which corresponds to a local averaging:
\bea
\sigma^{LMA}_{ij}[w(\br)] &=& \frac{1}{N_{\text{site}}} \sum_{\br_i} \sigma_{ij, U} (w(\br_i)),
\label{eq:lma_simga}
\eea
where $\sigma_{ij, U}$ is the set of easily calculable homogeneous Kubo formula conductivity tensors assuming that the local Weiss field $w(\br_i)$ is uniformly 
applied across the entire system.

\begin{figure}
\centering
\includegraphics[width= 0.9\textwidth]{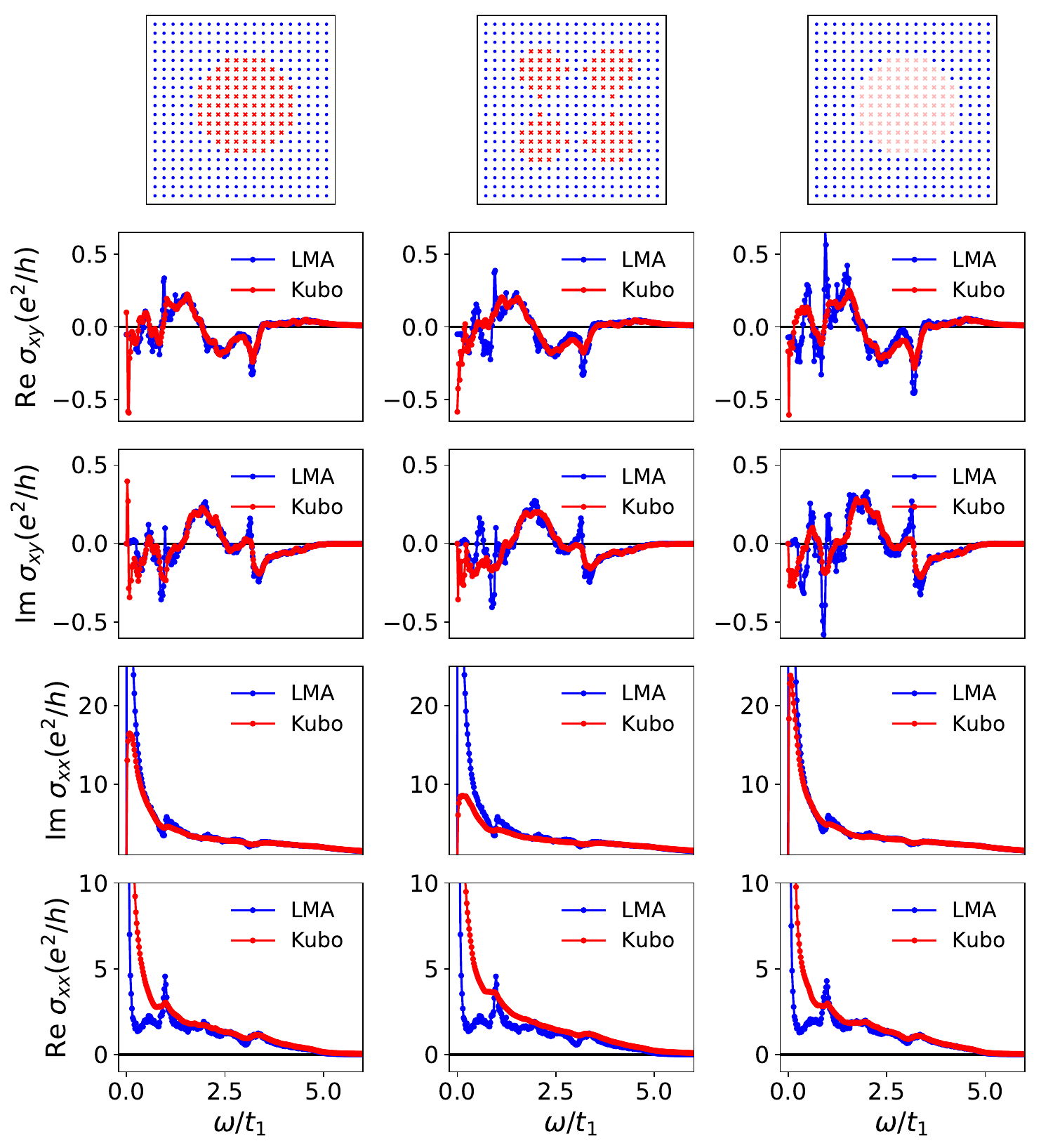}
	\caption{Each column shows (1) the unit cell of a super-cell Weiss-field profile $w(\br_i)$, whose dimension is given by $N_x \times N_y \times 1 = 30 \times 30 \times 1$ and (2) the corresponding conductivity versus frequency. The bottom four panels show the comparisons between the exact results obtained from the Kubo formula and those from LMA for components of the conductivity tensor. We observe that LMA works well at high frequency (e.g. $\omega/t_1 > 2$), in line with the semiclassical picture described in the main text. Here, we have used $t_2/t_1 = -0.2, f/t_1 = 0.4, \lambda_{\rm soc}/t_1 = 0.4$. In the unit cells, dots (blue) and crosses (red)
	correspond to the exchange splitting Weiss field 
	$(w_{\text{dot}}, w_{\text{cross}}) = (1.5 t_1, -1.5 t_1), (1.5 t_1, -1.5 t_1)$ and $(1.5 t_1, -0.75 t_1)$ respectively from left to right.}
	\label{fig:KuboLMA}
\end{figure}

Figure \ref{fig:KuboLMA} presents the comparison between the exact Kubo formula super-cell calculation and the LMA for different inhomogeneous Weiss field configurations.
The topmost row shows the unit cells of three super-cell profiles. Dots (blue) and crosses (red) denote the positive and negative z-direction of the Weiss field respectively, while
the intensity of the color denotes the magnitude of the Weiss field. Below each configuration, we plot the corresponding conductivity tensor, showing its different components.
We observe that LMA results (blue) and the exact results (red) agree well with each other in the large frequency limit, e.g. $\omega > 2t_1$, for all the tensor components. 
This corresponds to $\omega \gtrsim W/4$, where the bandwidth $W \sim 8 t_1$.
We have also found good agreement for other sets of model parameters and electronic densities, so the LMA appears to be a robust approximation for the optical
conductivity tensor.

Turning to the polar Kerr angle $\theta_K$, this is related to the dielectric functions via \cite{kerrangle}
\bea
\theta_K &=& \Re\left(\frac{\varepsilon_{xy}}{(\varepsilon_{xx}-1)\varepsilon_{xx}^{1/2}}\right),
\label{eq:moke}
\eea
where the dielectric function depends on the conductivity through the following relation $\varepsilon_{ij} = \varepsilon_b ~\delta_{ij} + \I\frac{\sigma_{ij}}{\omega\varepsilon_0}$. $\varepsilon_b$ is the background dielectric, and $\varepsilon_0$ is the vacuum permittivity. 
For SrRuO$_3$ in the optical frequency regime, the denominator, which involves only the longitudinal component $\varepsilon_{xx}$, appears less 
sensitive to the temperature or the magnetization (e.g., it
exhibits no sign changes), and hence less sensitive to the Weiss field. This has been reported in Ref. \cite{temperature_insensitive} (see also the flat temperature 
dependence in the optical regime of an empirical expression for $\sigma_{xx}(\omega)$ in Ref. \cite{empirical_temperature_insensitive}). Thus $\theta_K$ is
closely tied to $\sigma_{xy}(\omega)$, and we may then directly apply the LMA to the Kerr angle as in Eq. (\ref{eq:LMA_kerr}). 
(In our calculations above, we have assumed that the magnetization is linearly proportional to the Weiss field, which we have checked is correct for the 
regime of interest). The LMA is thus a useful approximation for studying Kerr angle in the optical regime in the presence of magnetic domains 
as discussed in the main text.

\clearpage

\section{S6. A\lowercase{nsatz for the two-domain model and magnetization fits}}
\label{ansatze_construction}
The two-domain model is a simplified description for the domain evolution during the magnetization reversal near the coercive magnetic field. As described in the main text, the model characterizes a magnetization profile by replacing the positive domains having positive z-component magnetization with an effective value $\hat{z} M_+$ while replacing the negative domains with an effective value $-\hat{z} M_-$. The volume fraction of the positive domain is denoted by $x_+$, so the negative domain has a volume fraction $x_- = 1 - x_+$.

Consider an increasing-field sweep starting from a sufficiently large negative field $-B^*\hat{z}$ to a large positive field $+B^*\hat{z}$ during which the magnetization begins from a uniform value $-M_{sat}\hat{z}$ and eventually reaches a uniform value $+M_{sat}\hat{z}$, passing through a magnetization reversal region where minority
magnetic domains grow and proliferate.
For our data, we can set $B^* \sim 0.5$ Tesla.
We can conveniently divide the field range into two regimes: (1) a negatively uniform regime $B < 0$ where $x_+ = 1 - x_- = 0$, and (2) 
a magnetization-reversal regime $B > 0$ where $x_+$ increases towards $1$ while $x_-$ drops to zero.
In the uniform regime with $B < 0$, the field dependence of $M_-$ is directly given by the experimentally measured magnetization $M(B)$, namely $M_-(B) = - M(B)$ 
in regime (1). In the rest of this section, we construct ansatz for $x_{\pm}(B)$ and $M_{\pm}(B)$ in the regime $B > 0$.

The ansatz for $x_+(B)$ in an increasing-field sweep must satisfy two constraints. First, it must obey boundary conditions: $x_+(0) = 0$ and $x_+(B^*) = 1$. Second,
in the presence of the positive applied field, it should be monotonically increasing with $B$. We assume that $x_+(B)$ follows a modified hyperbolic tangent function (shifted and rescaled), as given below.
\bea
x_+(B) &=& \left[ \frac{\tanh \left(\frac{B - b}{w}\right) + \tanh\left(\frac{b}{w}\right)}{\tanh \left(\frac{B^* - b}{w}\right) + \tanh\left(\frac{b}{w}\right)}\right] \left(\frac{B}{B^*}\right)^{\delta} \equiv F(B,b,w),
\eea 
The parameters $b$ and $w$ 
determine the behaviour of $F$. Physically, $b$ is roughly set by the coercive field, while a smaller $w$ results in a more rapid change of $x_+$ as we go across $B=b$.
The envelope function $(B/B^*)^{\delta}$ with $0<\delta < 1$ is needed to account for a slow increase of the magnetization near the end of the magnetization-reversal regime
approaching $B=B^*$.

\begin{figure}[b]
\centering
\includegraphics[width = 0.99\textwidth]{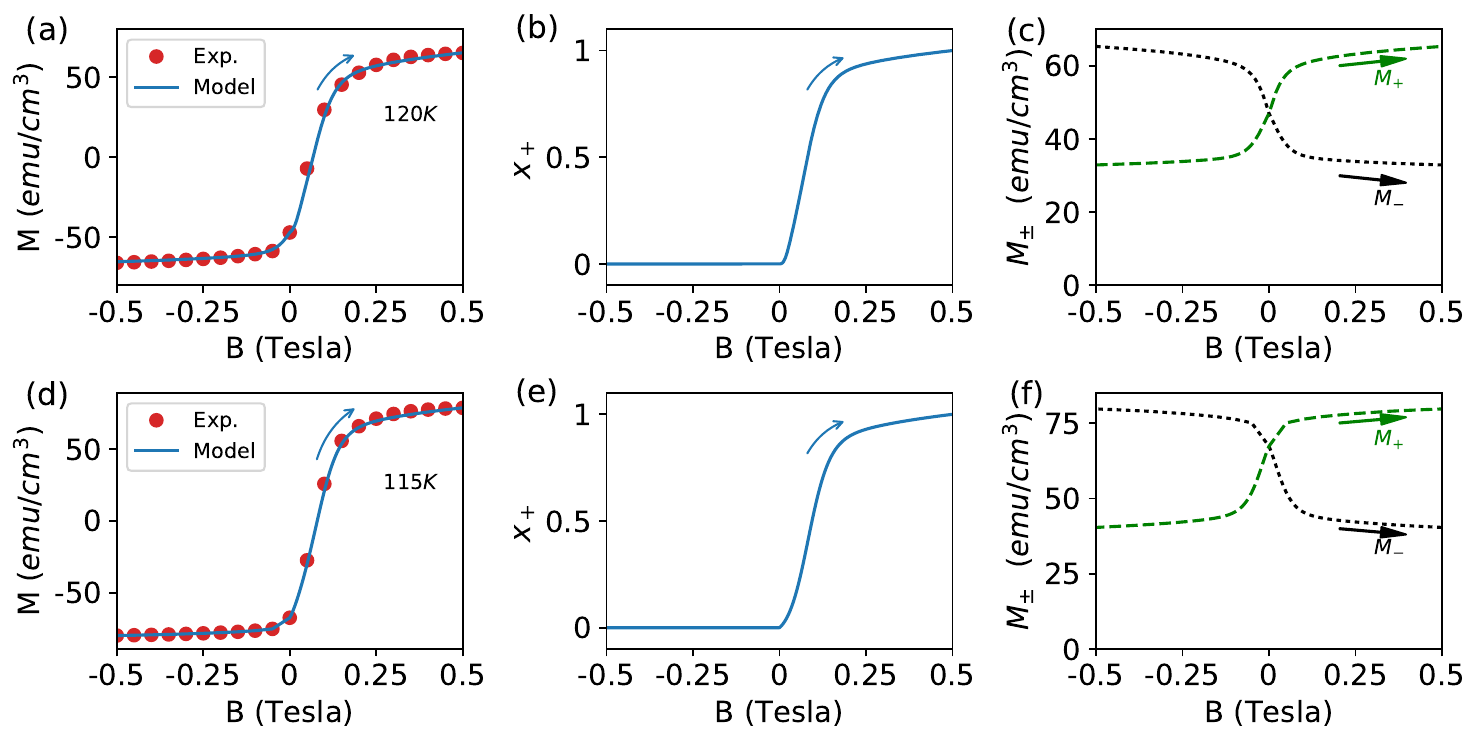}
\caption{Ansatz $x_+$ and $M_{\pm}$ obtained from fitting to the experimental magnetization curves (first column) at 120 K for (a)-(c) and 115 K for (d)-(f) in the increasing-field sweep direction. These are then used to compute the Kerr angle shown in the main text.}
\label{fig:lma_params}
\end{figure}

Turning next to the ansatz for $M_\pm (B > 0)$, we will use time-reversal symmetry to set $M_+(B > 0) = M_-(-B) = -M(-B)$, 
which is thus fully determined from the experiment. Furthermore, $M_+(B^*)=M_{sat}$ since $x_+=1$ at this field.
The impact of a positive field on $M_-$ is unknown. We assume that the field dependence of $M_-(B\geq 0)$ has a similar tanh form,
decreasing from the experimentally measured $-M(0)$ to a value smaller in magnitude $- \zeta M(0)$ at $B^*$, where $0 < \zeta < 1$. This can be achieved by the following function:
\bea
M_-(B) &=& \left[1 - (1 - \zeta)~ F(B,b',w')\right] M_0,
\eea  
where we use the same simple tanh functional form as before, but with different fit parameters $b'$ and $w'$. 
We determine the ansatz parameters $(b,w,b',w',\delta)$ by 
fitting the average magnetization curve obtained from the model $M_{av}(B) = x_+(B)M_+(B) - (1 - x_+(B)) M_-(B)$ to the experimental curve. The results are then used to compute the Kerr angle: $\theta_K(B) = x_+(B)\theta_{K, U}(M_+(B)) + (1 - x_+(B)) \theta_{K, U}(-M_-(B))$, where $\theta_{K, U}(m)$ is the Kerr angle as a function of the uniform magnetization $m$ as defined in the main text. The result of the decreasing field sweep is then taken to be the time-reversal counterpart of that in the increasing-field sweep, namely $\theta_K(B) \rightarrow -\theta_K(-B)$. Figure \ref{fig:lma_params} illustrates two examples of the fitting, and the ansatz corresponding to the Kerr angle plots at 115K and 120K in the main text.

\clearpage

\section{S7. I\lowercase{mpact of modifying the shape of the} $\theta_{K, U}$ \lowercase{curves}}
\begin{figure}[b]
\centering
\includegraphics[width=\textwidth]{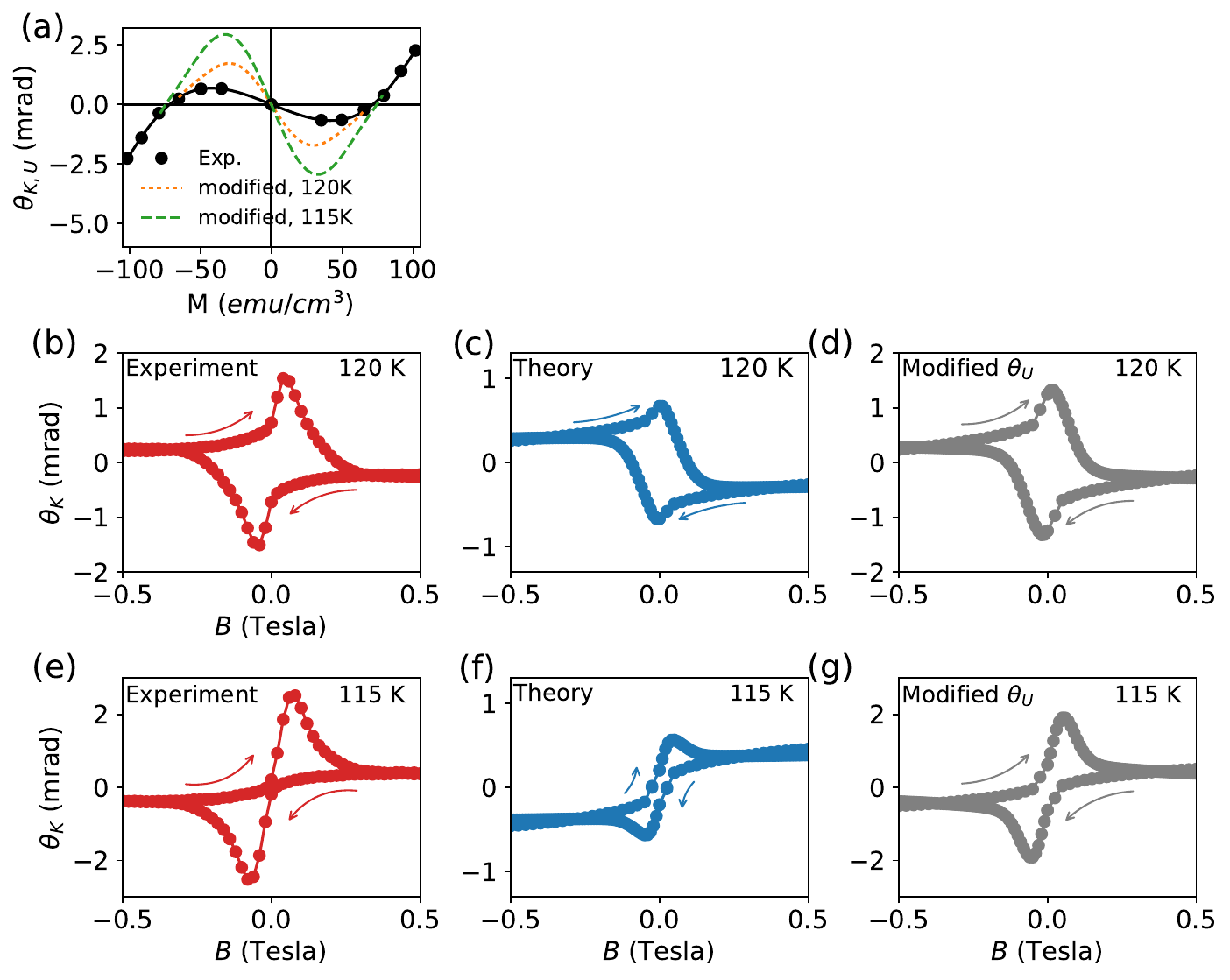}
\caption{(a) $\theta_{K, U}$ curves obtained from the experiment (dots) and obtained from its slight modifications. Each modified curve must terminate at the experimental data point for that corresponding temperature. (b) and (e) Experimental data at 120K and 115K. (c) and (f) Kerr angle obtained from LMA with the experimental $\theta_{K, U}$ curve from panel (a). 
(d) and (g) Kerr angle obtained using the modified $\theta_{K, U}$ curves in panel (a).}
\label{fig:modified}
\end{figure}

An essential input to our phenomenological theory is the magnetization dependence of the Kerr angle in a uniform magnetization profile, i.e., $\theta_{K, U}(M)$. In the main text, we extract
this from the experiment, by measuring the Kerr angle in a large applied field, while tuning $M$ via temperature. This leads to a good 
qualitative agreement between the theory and the experiment. However, the model in fact requires as input an {\it isothermal} $\theta_{K, U}(M)$ curve, 
which is beyond the scope of our current study. 
Instead, we show in Fig. \ref{fig:modified} how assuming that the isothermal curve deviates slightly from the
experimental $\theta_{K, U}$ curve, can lead to a better quantitative agreement with the measured Kerr anomaly. The proposed isothermal curves
are indicated by the dashed lines in panel (a), which  terminate at the experimental data points at the corresponding temperature (e.g., the proposed $T=115$\,K green curve 
touches the experimental black curve at $115$\,K).

\clearpage

\section{S8. I\lowercase{mpact of domain walls on dc \uppercase{H}all conductivity}}
\label{domain_wall}

\begin{figure}[b]
\centering
\includegraphics[width=0.7\textwidth]{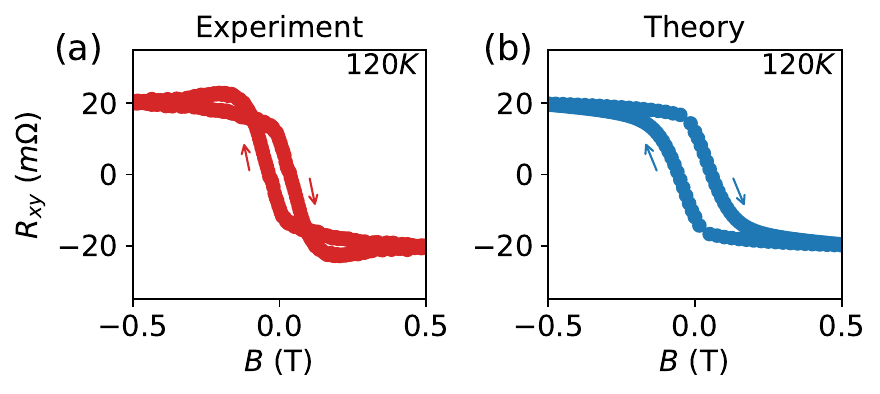}
\caption{(a) Hall anomalies in the 88 nm SrRuO$_3$ sample at 120 K. (b) A Hall-resistance hysteresis loop obtained from an effective medium approximation, showing no obvious anomalies. In the increasing field sweep, a negative correction is needed to account for the bump-like feature seen in the experiment. We discuss in the text that such an
anomalous correction may arise when we take magnetic domain walls into account.}
\label{fig:LMA_rxy}
\end{figure}

In this section, we show from a numerical computation using the Kubo formula that the dc Hall conductivity exhibits a domain wall contribution
$\sigma_{xy}^{\text{DW}}$, which may explain the Hall anomalies observed during magnetization reversal in our experiment.

We first compute the Hall resistivity hysteresis loop using an effective medium approximation (EMA) \cite{EMA1}, which is a well known
approximation for computing dc transport coefficients in an inhomogeneous system. Since dc transport is non-local, the EMA may be viewed as the
appropriate dc generalization of the LMA. Its formalism relies on solving Maxwell's equations with matching boundary conditions between constituents of the inhomogeneous system with well-defined local dc transport coefficients \cite{EMA1}. For simplicity, we deploy the spherical-inclusion version of EMA \cite{EMA2}.
Using the fitting functions $x_+(B)$ and $M_{\pm}(B)$ from the anomalous Kerr calculation and the curve of $R_{xy, U}$ versus magnetization from Fig. 1(f) in the main text, 
we can similarly compute the hysteresis loop for the Hall resistivity. The result is shown in Fig. \ref{fig:LMA_rxy} in comparison with the experimental data at 120 K. 
There are no obvious bump features in the EMA hysteresis loop. In the increasing-field sweep, a negative correction is needed to account for the anomalies.
Below, we show that there can be a domain-wall correction $\sigma_{xy}^{\text{DW}}$ arising from quantum mechanical effects beyond the EMA. 
We find that the domain wall contribution to the Hall conductivity has a positive sign for positive magnetization,
which implies a negative correction to the Hall resistivity $\rho_{xy} \approx -\sigma_{xy}/\sigma_{xx}^2$ in an increasing field sweep.
This domain wall contribution, which is expected to become more important near the magnetization reversal, 
may thus account for the observed anomalous dip in $R_{xy}$ in the increasing field sweep.

\begin{figure}
\centering
\includegraphics[width=0.8\textwidth]{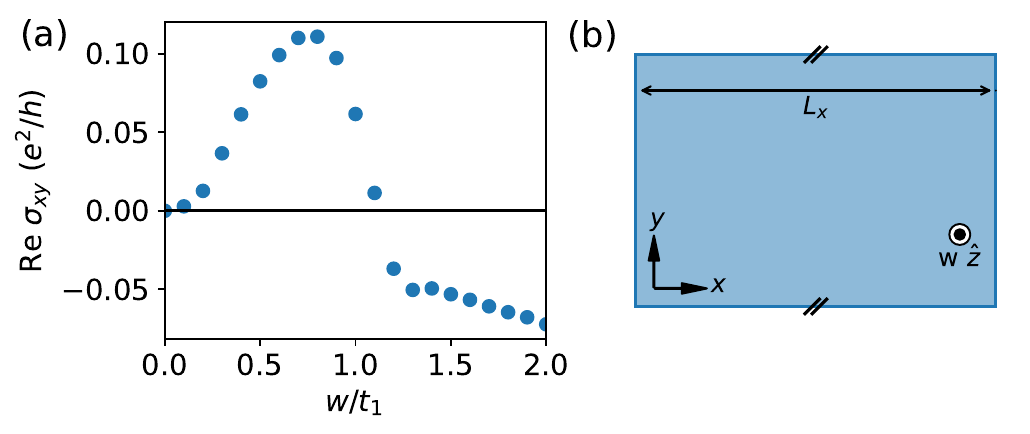}
\caption{(a) dc Hall conductivity $\sigma_{xy}$ as a function of the exchange-splitting Weiss field $w$, featuring a nonmonotonic behaviour of $\sigma_{xy}$ similar to the nonmonotonic magnetization dependence of the Hall resistivity observed in the experiment (see also \cite{Fang2003}). Our result is obtained from the cubic $t_{2g}$ model on a geometry with open boundaries in the x-direction and periodic boundary conditions in the y and z-direction as shown in (b). The Weiss field is uniform, and the dimension of the system is given by $L_x = 60, L_y = 215, L_z = 215.$ We have used the hopping parameters: $t_1 = 1, t_2 = 0.3, f = 0.2, \lambda_{soc} = 0.25$.}
\label{fig:appdx_uniform}
\end{figure}

\begin{figure}
\centering
\includegraphics[width=0.95 \textwidth]{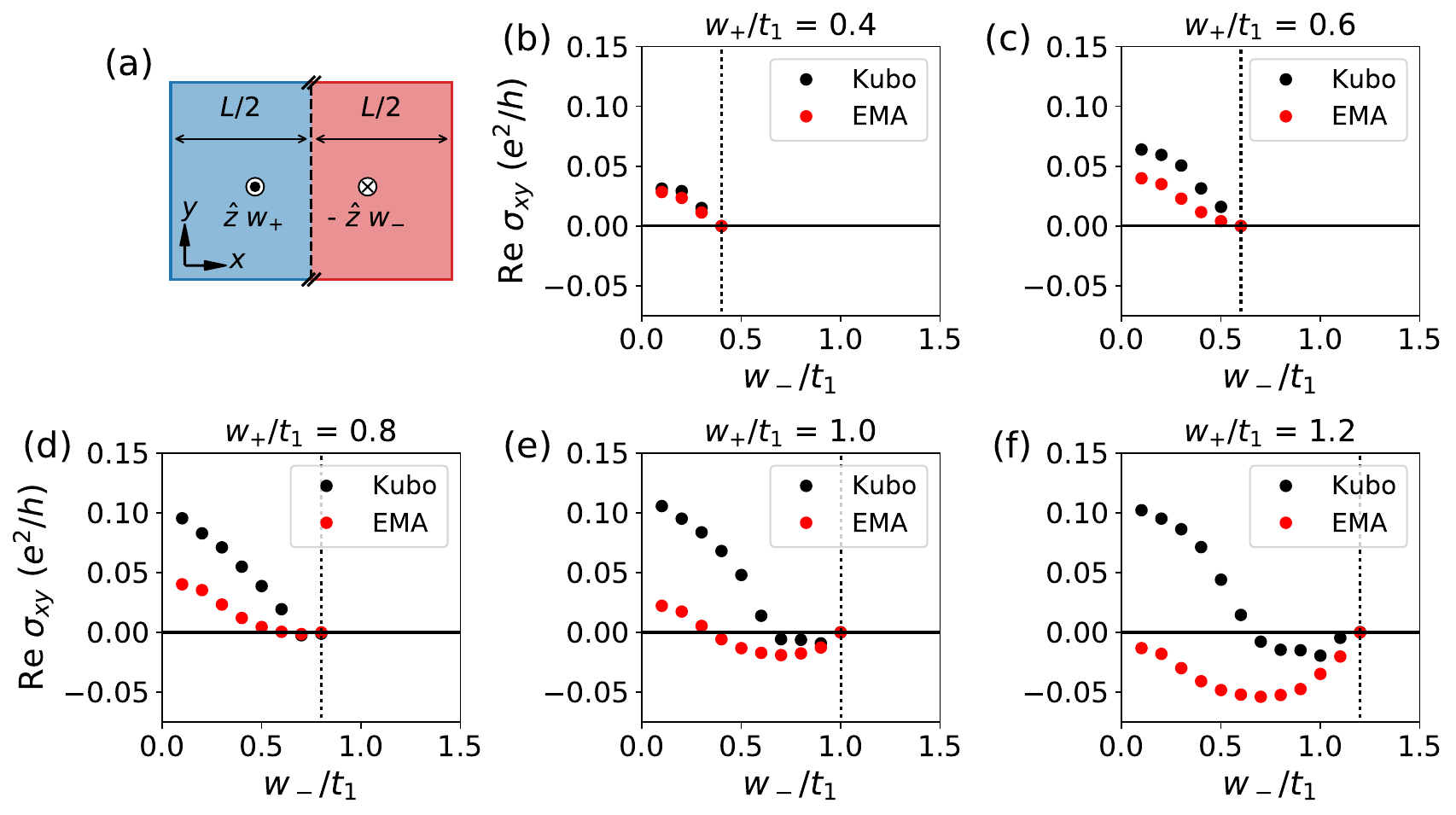}
\caption{(a) Weiss-field profile with a sharp domain boundary in the yz-plane which divides the system into two equal domains with different Weiss fields: $\pm w_{\pm}\hat{z}$. (b)-(f) dc Hall conductivity $\sigma_{xy}$ as a function of $w_-$ for various values of $w_+$, obtained from Kubo formulae versus that obtained from EMA. The correction from the domain wall $\sigma_{xy}^{\text{DW}} = \sigma_{xy}^{\text{Kubo}} - \sigma^{\text{EMA}}$ is clearly visible. The dashed lines mark the places where $w_+ = w_-$.
The dimension of the system is identical to that in Fig. \ref{fig:appdx_uniform}.}
\label{fig:appdx_domain_wall}
\end{figure}

We consider the cubic-lattice $t_{2g}$ Hamiltonian, Eq. (\ref{eq:t2g_hamiltonian}), placed on a geometry with open boundaries in the x-direction and periodic boundary conditions in the y and z-direction, as illustrated in Fig. \ref{fig:appdx_domain_wall}(b). We choose the hopping parameters such that the Hall conductivity exhibits a nonmonotonic dependence on the Weiss field $w$, as shown in Fig. \ref{fig:appdx_uniform}(a), similar to that observed in the experiment (see $R_{xy, U}$ versus magnetization in Fig. 1 in the main text). We note that the sign of $\sigma_{xy}$ here is identified with that of $- R_{xy, U}$ in the experiment. We identify the Weiss field here with the magnetization, and we have checked numerically that they are indeed proportional to each other in the case with uniform Weiss-field configurations. This simple model is useful since it
qualitatively captures the sign of the Hall effect in SrRuO$_3$, and its change with magnetization, although we do not expect it to quantitatively explain the experimental data.

We next introduce a sharp domain wall in the yz-plane, as shown in Fig. \ref{fig:appdx_domain_wall}(a), which divides the system into two equal domains with two Weiss fields: $\pm w_{\pm}\hat{z}$. We consider the regime where $w_+ > w_-$, following our fits to the magnetization data in the regime $B > 0$ of the increasing field sweep. 
We then compute $\sigma_{xy}$ using two methods: (1) exact Kubo formula and (2) the effective medium approximation. In the latter, we use the response functions from the uniform calculation done in Fig. \ref{fig:appdx_uniform} as the inputs. 
We identify the difference between the result from these two calculations as the domain wall contribution to the Hall conductivity:
$\sigma_{xy}^{\text{DW}} = \sigma_{xy}^{\text{Kubo}} - \sigma_{xy}^{\text{EMA}}$. 

Figure \ref{fig:appdx_domain_wall}(b)-(f) shows the Hall conductivity in the $w_- < w_+$ regime for various values of $w_+$. We find that the Kubo formula result lies above the
EMA result, and we thus infer a positive the domain-wall correction $\sigma_{xy}^{\text{DW}}$. This is precisely the correct sign needed to account for the Hall anomalies.
In literature, the nonmonotonicity in the Hall effect with varying the uniform magnetization
has been shown to arise from topological Weyl points in the band structure \cite{Burkov2013}, 
so it is possible that the Hall anomaly may be the result of an interplay 
between the Weyl points and the magnetic domain walls. Indeed, domain walls have been shown
in Refs. \onlinecite{BalentsPRL2017Mn3X,BehniaNComm2019Mn3Sn} to enhance the Hall effect
for antiferromagnetic Weyl metals Mn$_3$Ge and Mn$_3$Sn. We will discuss this interplay for a model ferromagnetic Weyl metal in a future publication.

\bibliography{arxiv_version}

\end{document}